\documentclass[]{JHEP3}

\JHEPspecialurl{http://jhep.sissa.it/JOURNAL/JHEP3.tar.gz}

\usepackage{epsfig,multicol,bbm}
\usepackage{axodraw}
\newcommand\query[1]

\newcommand{\mrm}[1]{\mathrm{#1}}

\newcommand{\br}[1]{\overline{#1}}

\newcommand{\pT}{p_{\perp}}
\renewcommand{\d}{\mathrm{d}}

\newcommand{\be}{\begin{equation}}
\newcommand{\ee}{\end{equation}}
\newcommand{\ba}{\begin{eqnarray}}
\newcommand{\ea}{\end{eqnarray}}

\title{Visible Effects of Invisible Hidden Valley Radiation}
\author{Lisa Carloni  and  Torbj\"{o}rn Sj\"{o}strand \thanks{Work supported by Marie-Curie Early Stage Training program "HEP-EST'' (contract number MEST-CT-2005-019626) and partly by the Marie-Curie MCnet program.}\\ 
Theoretical High Energy Physics\\
Department of Astronomy and Theoretical Physics, Lund University,\\
S\"olvegatan 14A, SE 223-62 Lund, Sweden\\
E-mail: \email{lisa.carloni@thep.lu.se} and \email{torbjorn@thep.lu.se}
}

\preprint{
LU TP 10-17\\
MCnet/10/11
}

\abstract{
Assuming there is a new gauge group in a Hidden Valley, and a new type of
radiation, can we observe it through its effect on the kinematic distributions
of recoiling visible particles? Specifically, what are the collider signatures
of radiation in a hidden sector? We address these questions using  a generic
$SU(N)$-like Hidden Valley model that we implement in \textsc{Pythia}.  We
find that in both the $e^+e^-$ and the LHC cases the kinematic distributions
of the visible particles can be significantly affected by the valley
radiation. Without a proper understanding of such effects, inferred masses 
of ``communicators'' and of invisible particles can be substantially off.
}


\begin{document}

\section{Introduction}

%where  we discuss how to distinguish between discrete, global and gauge symmetries
One common feature in New Physics models is the conservation (or near conservation) of a new quantum number. Often it is associated with a parity symmetry, like R-parity in supersymmetric models or T-parity in Little Higgs ones. Such conserved parity-like symmetries serve two basic model-building purposes: firstly, they forbid odd-parity  tree level corrections to electroweak precision observables, and secondly, they make the lowest lying odd-parity state stable, thus providing a possible dark matter candidate. 
The new charge may alternatively come from a continuous symmetry, a global  symmetry  or a new gauge symmetry, for example, and still fulfill the same purposes. 

Regardless of the specific model realization, we can imagine that a  new conserved quantum number is discovered at LHC. 

In this article, we wish to take some first steps towards addressing a general 
phenomenological question:
\emph{if a new apparently conserved charge should be discovered, is it possible to
 determine experimentally whether it arises 
from a discrete, a global or a gauge symmetry?
Specifically, is it possible to determine 
whether it is the source of a new field?}

In principle, a continuous symmetry has additive quantum number conservation whereas a discrete one has multiplicative conservation. To distinguish between gauge and global symmetries one could look for Gauge bosons, for Goldstone bosons and in general at the particle spectrum. 
The new sector however may be "hidden''. That is, the carriers of the new symmetry, for the two basic reasons mentioned above, could lie entirely within the new sector and be neutral  under (or have very weak) SM interactions. Indeed, a new unbroken symmetry would have to be invisible or else it would have already been found. Thus any radiation or other dynamic phenomena 
associated with it would be invisible to SM matter. 

If the charge did radiate in the new sector, would we be still be able to observe indirectly 
the effects of the hidden radiation? How would the kinematic distributions of the visible 
particles be affected? Could we extract information from these kinematic distributions 
 about the dynamics within the hidden sector? Could one distinguish Abelian from non-Abelian 
gauge groups, study the different particle (or unparticle) contents or measure the strength of 
the couplings? This could lead to a better understanding of the higher-energy dynamics, 
the ultraviolet completion of the theory, the
symmetries involved, and possibly even the mechanism by which they are broken. 

%Hidden Valley scenario as ideal gym
The ideal terrain to begin to explore these effects is Hidden Valley models 
\cite{Strassler:2006im}. 
%where we explain what WE mean by Hidden Valley
We extend this name to the class of models satisfying 
the following criteria. 
First,  there must be a \emph{new light hidden sector} (the valley), 
decoupled from the visible SM one, that has not yet been discovered because of some barrier. 
This can be an energy barrier or of another nature, 
e.g.\ symmetry-forbidding tree-level couplings. 
Second, the decoupling of the new hidden valley sector from the visible SM one  
must happen at relatively low energies, around the TeV scale, 
in such a way that the cross sections for Standard visible particles disappearing into the 
hidden sector (and vice versa) are small enough to evade the current experimental limits, 
and yet large enough to be observable at  LHC. These experimental limits are of course model dependent, as we will discuss in sections \ref{sec:HVscenarios} and \ref{sec:SU(3)}.
  
Typically, the valley particles "v-particles'' are charged under a valley group $G_v$ and neutral under the SM group $G_{SM}$, and the SM particles are neutral under $G_v$. In order to have interactions between the two sectors there has to be a "communicator'' which couples to both SM and valley particles. A common choice is to have a coupling via  a $Z^{\prime}$ or via loops of heavy particles carrying both $G_{SM}$ and $G_v$ charges.
 
Examples of Hidden Valleys can be found in many models, such as String Theory \cite{Blumenhagen:2005mu}, Twin Higgs models \cite{Chacko:2005pe}, folded SUSY \cite{Strassler:2006qa,Zurek:2008qg}, and Unparticle models \cite{Georgi:2007ek,Strassler:2008bv}.

Hidden Valley scenarios can naturally provide  candidates for Dark Matter and 
can easily fit cosmological constraints. Just to give an example, in \cite{Strassler:2006im} 
the v-interactions ensure that all particles efficiently decay to the lightest
mesons. These mesons are allowed to annihilate to neutral $\pi^0_v$s, which
can then tunnel back into the SM. 
So long as the lifetime of the $\pi^0_v$ is  $\tau \ll 1$ sec, 
the number of $\pi$s left will decay exponentially 
before big-bang nucleosynthesis.

%where we explain why a Hidden Valley is an ideal scenario to study radiation
The reason why these scenarios are ideal to study the effects of radiation is the large disparity in the masses of the communicators and the v-particles. Typically, the communicator has a mass around the decoupling scale, say the  TeV scale, while the v-particle mass may be as low as 1--10 GeV. If both the communicator and the v-particle are charged under a new gauge $G_v$, they will radiate gauge bosons, and the larger their mass ratio the larger the amount of phase space available for the radiation, both in the normal and in the hidden sector. Thus if any effect at all are to be observable, it would be in this kind of scenario. 

%the hidden valley toy models 
%where we explain why a parton shower is the ideal tool to study is this case
We have devised a Hidden Valley  toy model to tackle the issue, and have implemented it in 
the \textsc{Pythia}~8  Monte Carlo event generator \cite{Sjostrand:2007gs}. The implementation allows for different valley flavour contents, particle masses, gauge groups and v-gauge couplings. In this way one may accommodate a range of different Hidden Valley scenarios. 

MC event generators offer flexible approaches to model radiation and parton shower evolution in great detail. 
One new central feature in the \textsc{Pythia}~8 implementation is the  ``competition'' mechanism 
between the hidden and the SM radiation, which is implemented as an ``interleaved''
shower, wherein different kinds of emissions, SM and hidden, can alternate if viewed in
terms of a common shower evolution scale. As a consequence, subsequent emissions 
in the visible sector, of gluons or photons, will then tend to have a lower energy
than they would have had, had the hidden radiation not been there.   
This is the key mechanism whereby we gain access to the information about the radiation
 in the hidden sector.

The intention of this article is not a full-fledged experimental analysis of how a new sector 
should be discovered and explored, neither with respect to potential background
processes nor to detector-specific capabilities --- since our implementation is
publicly available, we safely leave it to the experimental community to assess.
What we want to ascertain here is if \emph{there are observable signals of hidden valley 
radiation at all}, at the simple parton and hadron levels. 

%where we explain why we look at MT2, and other boost-invariant observables
It is not trivial to decide which visible particle kinematic distributions one should study
to reveal valley radiation effects and  to discriminate between  different models.
For instance, at an $e^+e^-$ collider the rise of the communicator pair-production cross section 
near threshold could allow to determine its spin, 
and thereafter the absolute size of the  cross section could suggest the 
presence of new ``colour'' factors --- recall that the pair-production of 
particles in the fundamental representation of a new $SU(N)$ group gives 
a factor $N$ in the cross section. Such measurements would not directly 
probe the hidden sector, however: they would not reveal whether a new group 
is gauged, or what is the coupling strength in it. For a hadron collider,
like the LHC, the uncertainty in the event-by-event subcollision energy
$\sqrt{\hat{s}}$ undermines analyses solely based upon the value of cross-section. 
The best strategy  is thus to complement cross-section with 
invariant mass measurements and the study of other boost-invariant quantities  
(for a recent review see the proceedings \cite{Brooijmans:2010tn}). 

This is the reason why we choose to study MT2 
\cite{Lester:1999tx} distributions, which give relations between 
communicator and v-particle mass. These observables are specifically designed 
to be boost-invariant and to deal with BSM models in which more than one particle 
escapes detection, such as in our toy model. But we also study "hidden observables", like the 
invariant mass distribution of a hidden particle together with its associated hidden radiation.

%where we discuss the issue of LHC luminosity
The effects of the hidden radiation on these distributions and how much one may observe 
depends on the details of the scenario considered, of course, but  also depend heavily upon 
the collider type considered, on its center of mass energy, and on its integrated luminosity 
$\mathcal{L}$. 
We consider two different LHC scenarios, one for the early data 
(the first 18 to 24 months at $7$ TeV with
an expected integrated luminosity $\mathcal{L}=1$ fb$^{-1}$) and one for later
data ($\sqrt{s}=14$ TeV and integrated luminosity $\mathcal{L}=100$ fb$^{-1}$). 
The conclusions in the two cases will be quite different. For $e^+ e^-$
collisions we will mainly refer to an ILC at 800 GeV, though we mention CLIC production cross sections at 3 TeV.

The paper is organized as follows: we give a first overview of the various valley scenarios in the literature, in section \ref{sec:HVscenarios}, and then describe the tools that are now available in \textsc{Pythia} in section \ref{sec:MCtools}. In section \ref{sec:SU(3)} we explain the model and the main features of interleaved SM and hidden radiation. Finally, in the last two sections we study the effects of hidden radiation on collider phenomenology. In section \ref{sec:CLIC} we discuss the $e^+ e^-$ case  and in section \ref{sec:LHC} the LHC one. 

\section{Hidden Valley scenarios}
\label{sec:HVscenarios}

As mentioned in the introduction a Hidden Valley is a light hidden sector, consisting of  particles which, depending on the model, might have masses as low as
$10$ GeV. The detailed spectrum of the v-particles and their 
dynamics within the hidden valley  depends upon 
the valley gauge group $G_v$, the spin and number of particles present in the theory,
and the representation they belong to. 

The effects of the hidden sector on the visible particle spectra will depend upon the way the hidden sector communicates with the SM, whether it is via a Higgs, multiple Higgses, a $Z^{\prime}$, heavy sterile neutrinos or via loop of heavy particles charged under both SM and valley gauge interactions. 

We would like to give a panoramic view of the different Hidden Valley scenarios without going into details and to underline those features that may be simulated with the new tools. 

%where we discuss the first hidden valley scenario
The simplest possibility is a QCD-like scenario, with a strong coupling constant, which may run like the QCD coupling does, with QCD-like hadronization generating valley pions, v-$\eta$s, v-$K$s, v-nucleons etc. The Standard Model $SU(3)_c\times SU(2)\times U(1)$ sector could couple ultra-weakly with the hidden $SU(N)$ sector via a neutral $Z^{\prime}$. %the coupling being ultra weak because of a small  $Z^{\prime}\bar{f}f$ interaction with the SM fermions, or because  of a small $Z^{\prime}$ coupling to the new sector, because of the large mass of the $Z^{\prime}$ or all of the above. 
%where we explain that these scenarios have stirred some interest
This scenario was investigated by Strassler and Zurek with tools analogous to
the ones used to simulate QCD \cite{Strassler:2006im}. It displays some rather startling features. For instance, a v-$\pi$ could have a displaced decay in the muon spectrometer in the ATLAS detector, resulting in a large number of charged hadrons traversing the spectrometer, or it could decay in the hadronic calorimeter  producing a jet with no energy deposited in the electromagnetic calorimeter and no associated tracks in the inner detector. Experimental studies for these scenarios are currently under way, by the D0, CDF, LHCb, ATLAS and CMS collaborations.

%where we introduce unparticles as hidden valleys and we show why a parton shower is needed
Typical hidden valley-like signatures appear also in Unparticle models with
mass gaps \cite{Strassler:2008bv}. These models display a conformal dynamic
above the mass gap, and a hidden valley behaviour when the conformal symmetry
is broken. Regardless of the dynamics above the mass gap, whether it is strongly coupled or weakly coupled, the signatures are similar to the ones mentioned in the previous scenario (displaced vertices and missing energy signals). This is because only the lower energy states, light stable hidden hadrons can decay back into Standard Model particles. The higher energy states, be they narrow resonances or a continuum of resonances, decay rapidly to these lower light stable hadrons.
As for the previous scenario, a parton shower is a key tool to study these models, not so much to determine the  phenomenology qualitatively, but  because it the only element of the hidden dynamics which is sensitive to the higher-energy conformal (or next-to-conformal) dynamics. The conformal dynamics will be reflected in the parton shower evolution, which can be rather different from the regular QCD one, especially in theories with a strong dynamics above  the mass gap.

There are of course many other Hidden-Valley related models, such as Quirky models  \cite{Kang:2008ea}, just to give an example, in which the parton shower evolution does not play the key role it does in the previous cases. Typically their phenomenology is better captured in terms of string dynamics and string fragmentation.

%where explain that our model does not deal with string fragmentation but rather with the parton shower, so with the first two examples and not with the third class.

In this paper we do not address the issue of string fragmentation or
hadronization. Our main focus is on the parton shower, as this best captures the nature of the hidden radiation.   

The model we built to investigate the existence of a this new radiation
exploits but a few features common in many hidden valley scenarios: the
presence of a new unbroken hidden gauge group, of a heavy communicator,
charged under both SM and hidden sector gauge group and decaying into a
visible and an invisible light particle, charged only under the new gauge
group. These characteristics fit many Hidden Valley models, we however make an
additional assumption, which is that the production cross sections\footnote{We
  will discuss the production cross sections in section
  \ref{sec:SU(3)}.} should be large enough for the effects of the hidden
radiation to be discernable. This model was then implemented in the \textsc{Pythia} event
generator. Notice however, that the shower mechanism we implemented is rather
different from the ones mentioned above, as we will discuss in the next
section.   

\section{Monte Carlo Tools in PYTHIA 8} 
\label{sec:MCtools}

In order to allow detailed studies of a set of scenarios, 
the models have been implemented in the \textsc{Pythia}
event generator, and will be publicly available from version 
8.140 onwards. 

\subsection{Particle content}

For simplicity we assume that the HV contains either an
Abelian $U(1)$ or a non-Abelian $SU(N_c)$ gauge group,
with spin 1 gauge bosons. The former group could be unbroken 
or broken, while the latter always is assumed unbroken. 
Casimir constants could be generalized to encompass other 
gauge groups, should the need arise, but for now we do not 
see that need. The gauge bosons are called $\gamma_v$ and 
$g_v$, respectively. 
  
A particle content has been introduced to mirror the 
Standard Model flavour structure. These particles, 
collectively called $F_v$, are charged under both the SM 
and the HV symmetry groups. Each new particle couples 
flavour-diagonally to a corresponding SM state, and has 
the same SM charge and colour, but in addition is in the 
fundamental representation of the HV colour, see 
Table~\ref{tab:particles}. Their masses and widths can
be set individually. It would also be possible to expand 
the decay tables to allow for flavour mixing.

\TABLE[t]{
%\renewcommand{\arraystretch}{1.15}
%\begin{center}
\begin{tabular}{|c|c|c||c|c|c|}
\hline
name & partner & code & name & partner & code \\
\hline
$D_v$ & $d$ & 4900001 & $E_v$ & $e$ & 4900011 \\
$U_v$ & $u$ & 4900002 & $\nu_{Ev}$ & $\nu_e$ & 4900012 \\
$S_v$ & $s$ & 4900003 & $MU_v$ & $\mu$ & 4900013 \\
$C_v$ & $c$ & 4900004 & $\nu_{MUv}$ & $\nu_{\mu}$ & 4900014 \\
$B_v$ & $b$ & 4900005 & $TAU_v$ & $\tau$ & 4900015 \\
$T_v$ & $t$ & 4900006 & $\nu_{TAUv}$ & $\nu_{\tau}$ & 4900016 \\
\hline
$g_v$ &    & 4900021 & & & \\
$\gamma_v$ & & 4900022 & & & \\
$q_v$ &    & 4900101 & & & \\
\hline
\end{tabular}
%\end{center}
\label{tab:particles}
\caption{The allowed particle content in the HV scenarios, 
with their SM partners, where relevant. The code is an integer 
identifier, in the spirit of the PDG codes, but is not part of 
the current Amsler:2008zzb standard.} 
}

These particles can decay to the corresponding SM particle,
plus an invisible, massive HV particle $q_v$, that then 
also has to be in the fundamental representation of the HV 
colour: $F_v \to f q_v$. The notation is intended to make  
contact with SM equivalents, but obviously it cannot be
pushed too far. For instance, not both $F_v$ and $q_v$ 
can be fermions. We allow the $F_v$ to have either of  
spin 0, $1/2$ and 1. Currently the choice of $q_v$ spin
is not important but, for the record, it is assumed to 
be spin $1/2$ if the $F_v$ is a boson and either of spin 0 
and 1 if $F_v$ is a fermion. 

\subsection{Production processes}

The HV particles have to be pair-produced. The production
processes we have implemented are the QCD ones, 
$g g \to Q_v \br{Q}_v$ and $q \br{q} \to Q_v \br{Q}_v$,
for the coloured subset $Q_v$ of $F_v$ states, and the 
electroweak $f \br{f} \to \gamma^*/Z^0 \to F_v \br{F}_v$ 
for all states. All of them would contribute at a hadron 
collider, but for a lepton one only the latter would be 
relevant. Each process can be switched on individually,
e.g. if one would like to simulate a scenario with only
the first $F_v$ generation.
   
Note that pair production cross sections contain a factor 
of $N_c$, with $N_c = 1$ for an $U(1)$ group, for the pair 
production of new particles in the fundamental representation 
of the HV gauge group, in addition to the ordinary colour 
factor for $Q_v$. Other things equal, this could be used 
to determine $N_c$ from data, as already discussed. For the 
case of a spin 1 $F_v$ it is possible to include an anomalous 
magnetic dipole moment, $\kappa \neq 1$. 

The spin structure of the $F_v \to f q_v$ decay is currently
not specified, so the decay is isotropic. Also the Yukawa
couplings in decays are not set as such, but are implicit in 
the choice of widths for the $F_v$ states.

The kinematics of the decay is strongly influenced by the 
$q_v$ mass. This mass is almost unconstrained, and can therefore
range from close to zero to close to the $F_v$ masses. We will
assume it is not heavier than them, however, so that we do 
not have to consider the phenomenology of stable $F_v$ particles. 

\subsection{Parton showers}

Both the $F_v$ and the $q_v$ can radiate, owing to their charge
under the new gauge group, i.e.\  $F_v \to F_v \gamma_v$ and
$q_v \to q_v \gamma_v$ for a $U(1)$ group, and $F_v \to F_v g_v$ 
and $q_v \to q_v g_v$ for a $SU(N_c)$ one. In the latter case 
also non-Abelian branchings $g_v \to g_v g_v$ are allowed.
Currently both $\gamma_v$ and $g_v$ are assumed massless, but
a broken $U(1)$ with a massive $\gamma_v$ is foreseen.

These showers form an integrated part of the standard final-state 
showering machinery. Specifically, HV radiation is interleaved
with SM radiation in a common sequence of decreasing $\pT$.
That is, at the stage before the $F_v$'s decay, they may radiate
$g$, $\gamma$ and $\gamma_v/g_v$, in any order. For the $i$'th 
emission, the $\pT$ evolution starts from the maximum scale
given by the previous emission. The overall starting scale 
$p_{\perp 0}$ is set by the scale of the hard process. Thus the 
probability to pick a given $\pT$ takes the form
\begin{equation}
\frac{\d \mathcal{P}}{\d \pT} = \left( 
\frac{\d \mathcal{P}_{\mrm{QCD}}}{\d \pT} + 
\frac{\d \mathcal{P}_{\mrm{QED}}}{\d \pT} +
\frac{\d \mathcal{P}_{\mrm{HV}}}{\d \pT}
\right) \;
\exp \left( - \int_{\pT}^{p_{\perp i-1}} \left(
\frac{\d \mathcal{P}_{\mrm{QCD}}}{\d \pT'} + 
\frac{\d \mathcal{P}_{\mrm{QED}}}{\d \pT'} + 
\frac{\d \mathcal{P}_{\mrm{HV}}}{\d \pT'} 
\right) \d \pT' \right)
\end{equation}
where the exponential corresponds to the Sudakov form factor.
Implicitly one must also sum over all partons that can radiate.

To be more precise, radiation is based on a dipole picture,
where it is a pair  of partons that collectively radiates a new
parton. The dipole assignment is worked out in the limit of 
infinitely many (HV or ordinary) colours, so that only planar
colour flows need be considered.
Technically the total radiation of the dipole is split 
into two ends, where one end acts as radiator and the other
as recoiler. The recoiler ensures that total energy and momentum
is conserved during the emission, with partons on the mass shell
before and after the emission. In general the dipoles will be 
different for QCD, QED and HV.

To take an example, consider $q \br{q} \to Q_v \br{Q}_v$,
which proceeds via an intermediate $s$-channel gluon. Since this 
gluon carries no QED or HV charge it follows that the 
$Q_v \br{Q}_v$ pair forms a dipole with respect to these two 
emission kinds. The gluon \textit{does} carry QCD octet charge,
however, so $Q_v \br{Q}_v$ do \textit{not} form a QCD dipole.
Instead each of them is attached to another parton, either
the beam remnant that carries the corresponding anticolour 
or some other parton emitted as part of the initial-state shower.
This means that QCD radiation can change the invariant mass of 
the $Q_v \br{Q}_v$ system, while QED and HV radiation could not.
When a $\gamma$ or $\gamma_v$ is emitted the dipole assignments 
are not modified, since these bosons do not carry away any charge.   
A $g$ or $g_v$ would, and so a new dipole would be formed. 
For QCD the dipole between $Q_v$ and one beam remnant, say, would
be split into one between the $Q_v$ and the $g$, and one further 
from the $g$ to the remnant. For HV the $Q_v \br{Q}_v$ dipole 
would be spit into two, $Q_v g_v$ and $g_v \br{Q}_v$. As the 
shower evolves, the three different kinds of dipoles will diverge
further.

Note that, in the full event-generation machinery, the final-state
radiation considered here is also interleaved in $\pT$ with the
initial-state showers and with multiple parton-parton interactions.

There is made a clean separation between radiation in the production
stage of the $F_v \br{F}_v$ pair and in their respective decay.
Strictly speaking this would only be valid when the $F_v$ width
is small, but that is the case that interests us here. 
In the decay $F_v \to f q_v$ the QCD and QED charges go with the 
$f$ and the HV one with $q_v$. For all three interactions the dipole is
formed between the $f$ and the $q_v$, so that radiation preserves 
the $F_v$ system mass, but in each case only the relevant dipole end is
allowed to radiate the kind of gauge bosons that goes with its charge.
(Strictly speaking dipoles are stretched between the $f$ or $q_v$ 
and the ``hole'' left behind by the decaying $F_v$. The situation 
is closely analogous to $t \to b W^+$ decays.)  

The HV shower only contains two parameters. The main one is the 
coupling strength $\alpha_v$, i.e. the equivalent of $\alpha_s$. 
This coupling is taken to be a constant, i.e. no running is included. 

 From a practical point of view it is doubtful that such a running 
could be pinned down anyway, and from a theory point of view it 
means we do not have to specify the full flavour structure of the 
hidden sector. The second parameter is the lower cutoff scale
for shower evolution, by default chosen the same as for the QCD 
shower, $p_{\perp\mathrm{min}} = 0.4$~GeV.

The HV showers are not matched onto higher-order matrix elements
for the emissions of hard $\gamma_v/g_v$ in the production process, 
and so contain an element of uncertainty in that region. For the
decay process the matching to first-order matrix elements have been
worked out for all the colour and spin combinations that occurs in
the MSSM \cite{Norrbin:2000uu}, and is recycled for the HV scenarios, with
spin 1 replaced by 0 for non-existing (in MSSM) combinations. This 
means that the full phase space is filled with (approximately) the 
correct rate. Some further approximations exists, e.g.\ in the 
handling of mass effects in the soft region. The chosen behaviour 
has been influenced by our experience with QCD, however, and so 
should provide a good first estimate. More than that we do not aim 
for in this study.

\section{The model: SM and $SU(3)_v$ radiation}
\label{sec:SU(3)}

To be specific, in the following we explore two similar Hidden Valley experimental scenarios. In the first,  the communicator $E_v$ is  a spin 1/2 particle charged under both the SM $SU(2)\times U(1)$ and  the valley gauge group $SU(3)_v$. We assume it has the same SM charges an electron would have, so it may be  pair-produced in $e^+e^-$ collisions, via  $Z/\gamma^{\star}$. Under the unbroken $SU(3)_v$, it transforms like a $\mathbf{3}$, so  it radiates both $\gamma$s  and massless hidden valley gluons $g_v$s. After the parton shower,  the $E_v$ eventually decays into a visible SM electron $e$ and an invisible spin 0 valley ``quark'' $q_v$. This $q_v$ belongs to the fundamental representation of $SU(3)_v$ and is not charged under the SM gauge group, so it only radiates $g_v$s.  See Fig. \ref{fig:EvEv}.     

The key feature is interleaved radiation, already introduced above. In the
current context it works as follows. Once the  $E_v$ has been produced it may
radiate a SM $\gamma$, say. This radiation will subtract energy from the $E_v$ and the following emission, be it another SM photon or a valley gluon, will have less phase space to radiate into. In an analogous way, assuming a valley $g_v$ is emitted next, it subtracts energy from $E_v$ and affects the following emissions which, again,  could be either visible or invisible. 

\FIGURE[ht]{
\epsfig{file=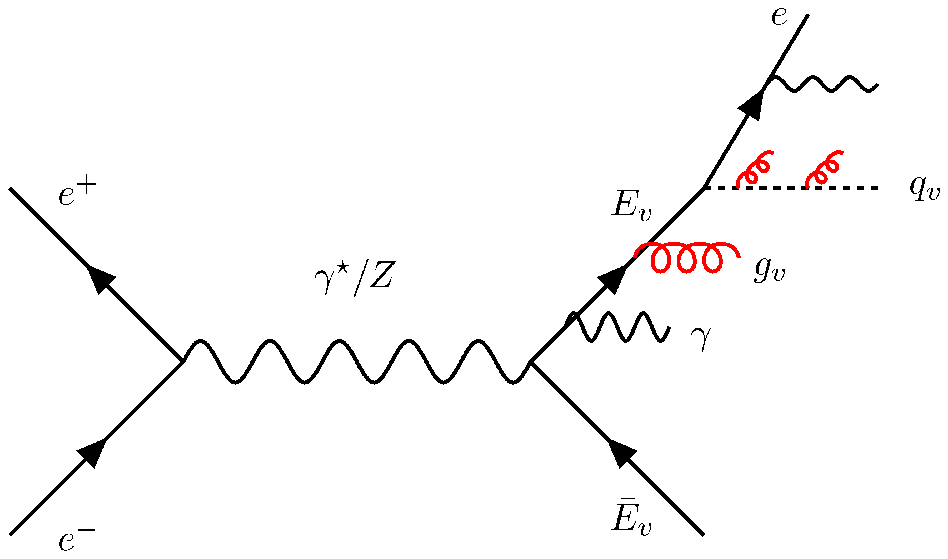,width=0.6\textwidth}
\caption{An $E_v \bar{E}_v$ pair is produced via $Z/\gamma^{\star}$. Since
  $E_v$ is charged under both $SU(2)\times U(1)$ and $SU(3)_v$, it radiates
  both $\gamma$s and $g_v$s. It eventually decays into $e$ and $q_v$. These
  then each radiate into their respective sector. Notice that $q_v$ here refers to a spin 0 particle.}
\label{fig:EvEv}
}

In the second scenario the communicator between SM sector and Hidden Valley
sector is a  quark-like, spin 1/2 object $Q_v$,  belonging to the
$(\mathbf{3},\mathbf{3})$ representation of the gauge group $SU(3)_c\times
SU(3)_v$.   The $Q_v$s are pair produced (mostly) via strong interactions
(gluon-gluon  or $q \bar{q}$ fusion). We choose the scenario in which only one
vector-like $Q_v$ is produced, the $D_v$. This $D_v$ emits massless valley
gluons (since the $SU(3)_v$ is assumed to be unbroken) and these may in turn
radiate more $g_v$s.  During the shower evolution,  \emph{both} types of
gluons are radiated until finally each $D_v$ decays into a visible SM $d$
quark an invisible spin 0 valley $q_v$. The decays are flavour diagonal, $D_v\rightarrow d+q_v$.

\FIGURE[ht]{
%\begin{minipage}[b]{0.45\linewidth}
\centering
\epsfig{file=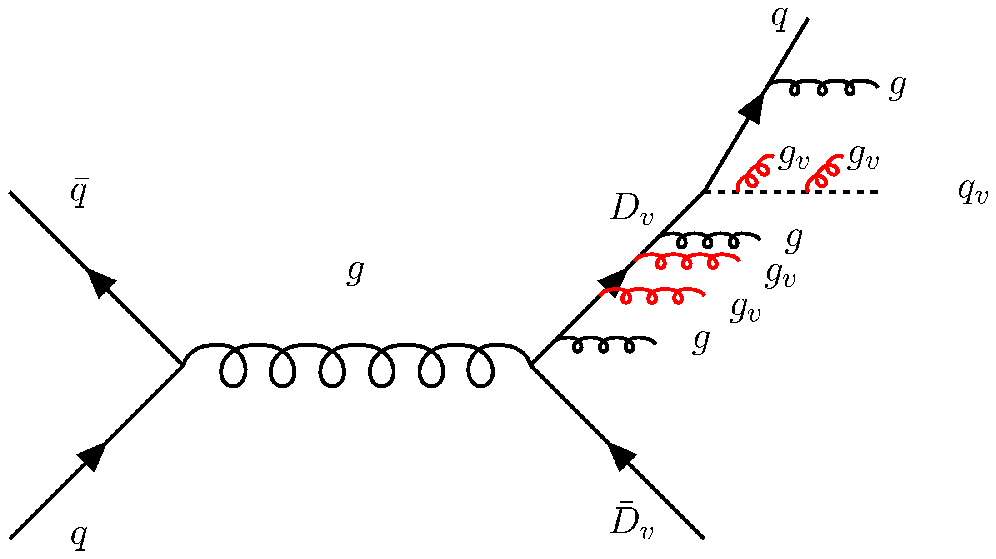,width=0.6\textwidth}
\caption{Pair production of hidden valley $D_v$s. Each $D_v$ can radiate $g$s
  and $g_v$s, and eventually decays $D_v\rightarrow q_v d$. The visible $d$
  then can radiate further $g$s and the invisible $q_v$ further $g_v$s.}
\label{fig:qq}
%\end{minipage}
}

The SM quark $d$ transforms as a $(\mathbf{3},\mathbf{1})$ under $SU(3)_c
\times SU(3)_v$, so it radiates only SM gluons,  while the valley $q_v$
belongs to the $(\mathbf{1},\mathbf{3})$ representation of $SU(3)_c \times
SU(3)_v$, so not having any SM color charge, it radiates only $g_v$s., see figure \ref{fig:qq}.

In both scenarios there are just three parameters left to vary:  the size of the valley coupling constant $\alpha_v$, the masses of the communicator particles $M_{E_v}$ or $M_{D_v}$ and the mass of the valley scalar $M_{q_v}$. 

Below, in Table \ref{tab:sigmas}, we list the total production cross sections
at different colliders: $e^+e^-$ with $\sqrt{s}=800$ GeV or $\sqrt{s}=3$ TeV,
and LHC with $\sqrt{s}=7$ TeV or $\sqrt{s}=14$ TeV for some typical $M_{E_v}, M_{D_v}$ mass values. 

\TABLE[t]{
%\renewcommand{\arraystretch}{1.15}
%\begin{center}
\begin{tabular}{|l|c|c||l|c|c|}
\hline
                                  & ILC            & CLIC     & &     LHC     &   LHC    \\
&(800 GeV)&(3 TeV)&&(7 TeV)&(14 TeV) \\
\hline
$M_{E_v}=300$ GeV   &              398    fb       &    44    fb        & $M_{D_v}=300$ GeV& $1.39\cdot 10^4$ fb &  $ 1.04\cdot 10^5$ fb \\
\hline
$M_{E_v}=500$ GeV   &                -                 &    41     fb       & $M_{D_v}=500$ GeV &         654     fb     &  $7.27\cdot 10^3$ fb \\
\hline
$M_{E_v}=1$ TeV       &               -                  &    32      fb       & $M_{D_v}=1$  TeV&        3.21    fb      &  $124$  fb\\
\hline
\end{tabular}
%\end{center}
\caption{The order of magnitude of the total production cross sections, in
  $fb$, at ILC (via $Z/\gamma^\star$), LHC (via $q \bar{q}$ or $gg$ fusion)
  with $\sqrt{s}=7$ TeV  and  $14$ TeV, for various values of the communicator
  mass. The spin of the communicator is assumed to be 1/2.}
\label{tab:sigmas}
}

We also show the spin dependence of the $E_v \bar{E}_v$ production cross
section at $e^+ e^-$ colliders for the three cases: $F_v$ spin 0 and $q_v$
spin 1/2,  $F_v$ spin 1/2 and $q_v$ spin 0 or 1, and $F_v$ spin 1 and $q_v$ spin 1/2, 
Fig.~\ref{fig:sigma_spin}.

\FIGURE[t]{\epsfig{file=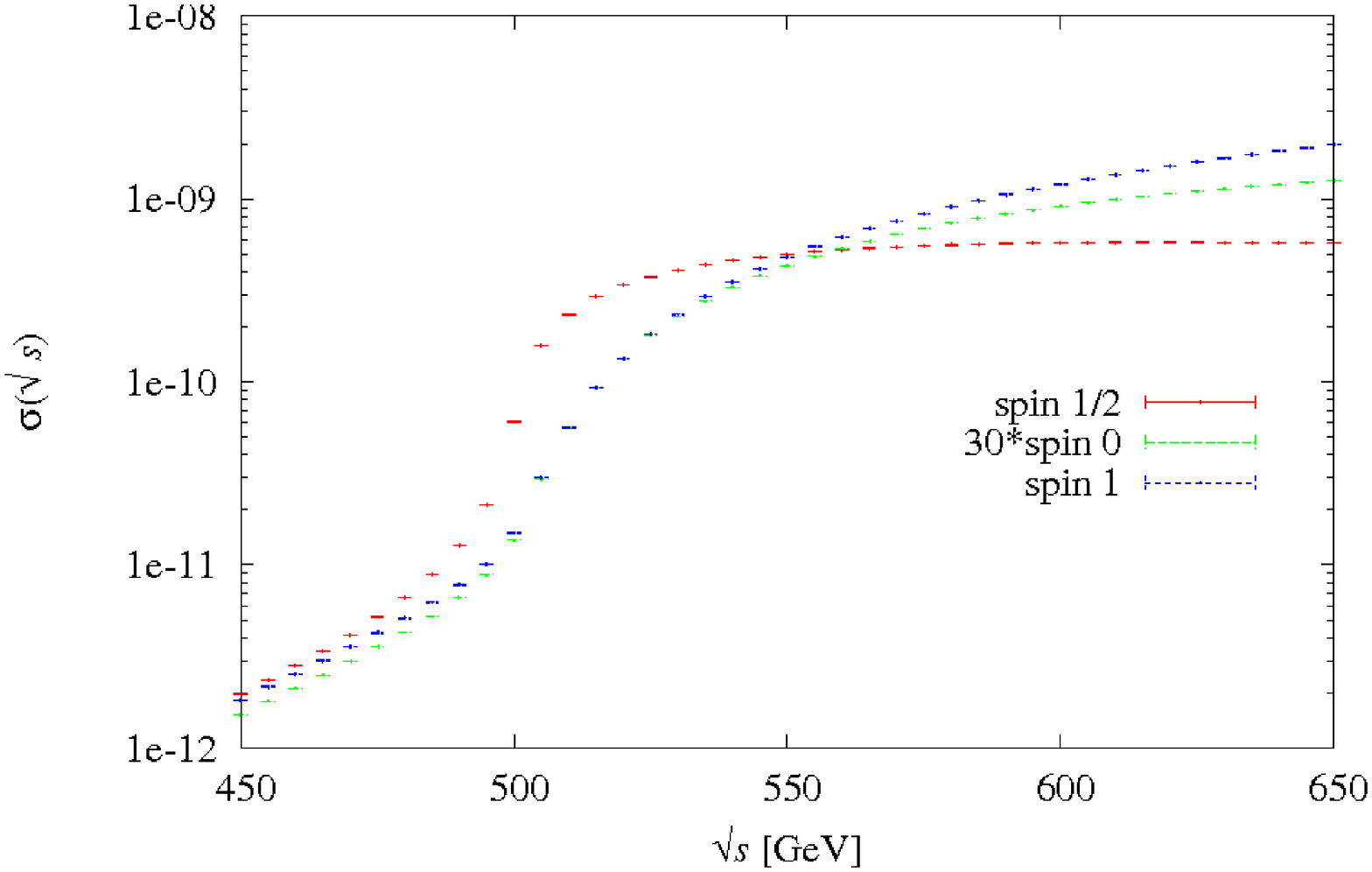,width=.4\textwidth,angle=270}
\caption{The spin dependence of the $E_v \bar{E}_v$ production cross section at $e^+ e^-$ colliders for $E_v$ spin 1/2, 0 and 1. $M_{E_v}=250$ GeV, the mass spread $\Gamma_{E_v}=2$ GeV, $M_{q_v}=50$ GeV.  The spin 0 curve has been scaled by a factor 30.}
\label{fig:sigma_spin}
}

The higher the spin, the larger the cross section. Indeed, the curve 
corresponding to $E_v$ spin 0 has been scaled by a factor 30 
to emphasize the similarity in shape with the spin 1 curve.
Note that the processes proceed through the $s$-channel exchange
of a spin 1 $\gamma^*/Z^*$. Thus the production of a spin $1/2$ pair 
has only a threshold factor $\beta$ from phase space, where $\beta$ 
is the velocity of the produced pair in the $\gamma^*/Z^*$ decay vertex,
while the other two have an (approximate) additional factor $\beta^2$ 
from helicity considerations. The results have again been obtained with 
a $SU(3)_v$ group, and are directly proportional to the $N_c$ chosen. 
Since we would not expect gauge groups with $N_c$ above (some multiple of) 
30, the conclusion would be that a threshold scan of the cross section
could be used to determine both the $E_v$ spin and the number of hidden
colours, as well as the $E_v$ mass, of course. A caveat would be that 
we have here only considered the $E_v$ gauge production mechanism,
not the possibility of a significant $t-$channel Yukawa contribution. The reason for not including this production channel is that it
would imply a large decay width for the $E_v$, which would give additional large and model dependent effects  to the cross section around threshold (see below). 

The experimental constraints on these two types of setup are similar to the ones for New Charged Leptons and Leptoquark production\footnote{We do not make use of the limits coming from the D0 or CDF  $\tilde{q}\rightarrow j+ E_T \hspace{-16pt}\not \hspace{10pt}$ searches, because these depend upon the chosen mSUGRA scenario.}.  
For the New Charged Leptons the PDG \cite{Amsler:2008zzb} gives the lower bound $m_{L^\pm} > 100.8$ GeV.  
For scalar and vector Leptoquark states we use the direct limits coming from leptoquark pair production and subsequent $LQ\rightarrow \nu q$ decay searches \cite{Abazov:2006wp}. For the all generation search $m_{LQ}>136$ GeV in the scalar case and $m_{LQ}>200$ GeV in the vector case, where one assumes the branching ratio $B(\nu q)=1$. Bounds on the leptoquark mass for the third generation decaying into $\nu b$ are more stringent \cite{Abazov:2007bsa}, $m_{LQ}>229$ GeV.  

We view these boundaries as simply indicative of the mass range we should contemplate and chose masses that are well beyond these boundaries. Our studies are in any case not critically dependent on them.

%the H1 collaboration \cite{Aktas:2005pr} gives $m_{LQ}\ge 276-304$ GeV. The ZEUS \cite{Chekanov:2003af} collaboration gives similar bounds.

We assumed the communicators to be massive, $M_{E_v}$ in the range [250,300]
GeV for the ILC case,  $M_{D_v}$ in the range [300, 500] GeV for the LHC 7 TeV
run, and [0.5,1] TeV for the 14 TeV run. The hidden scalar is taken to be light,
$M_{q_v}$ as light as 10 GeV. The $\alpha_v$ parameter is allowed to vary over
a wide range, and results shown for interesting values.
 
Depending on the size of these parameters, the effects of the radiation on the
lepton or quark kinematic distributions can be significant, as we will show in
the next two sections. But whether these effects will be observable strongly
depends on the statistics at hand. We assume  an integrated luminosity  
$L=200$ fb$^{-1}$ for the ILC, 1 fb$^{-1}$  for the 7 TeV LHC run and 
100 fb$^{-1}$ for the 14 TeV one. 

\FIGURE[t]{ 
\epsfig{file=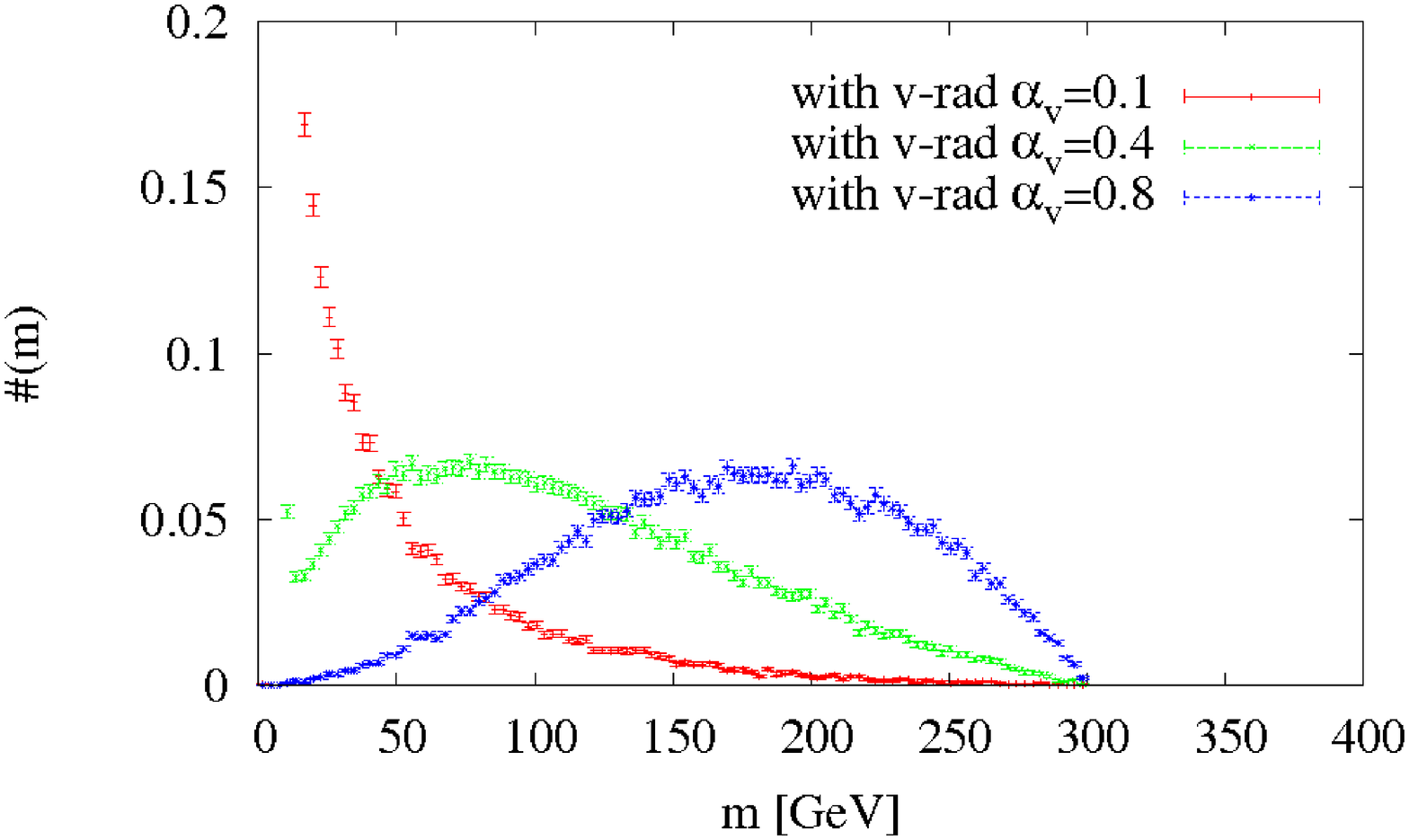, width=.315\textwidth, angle=270}
\epsfig{file=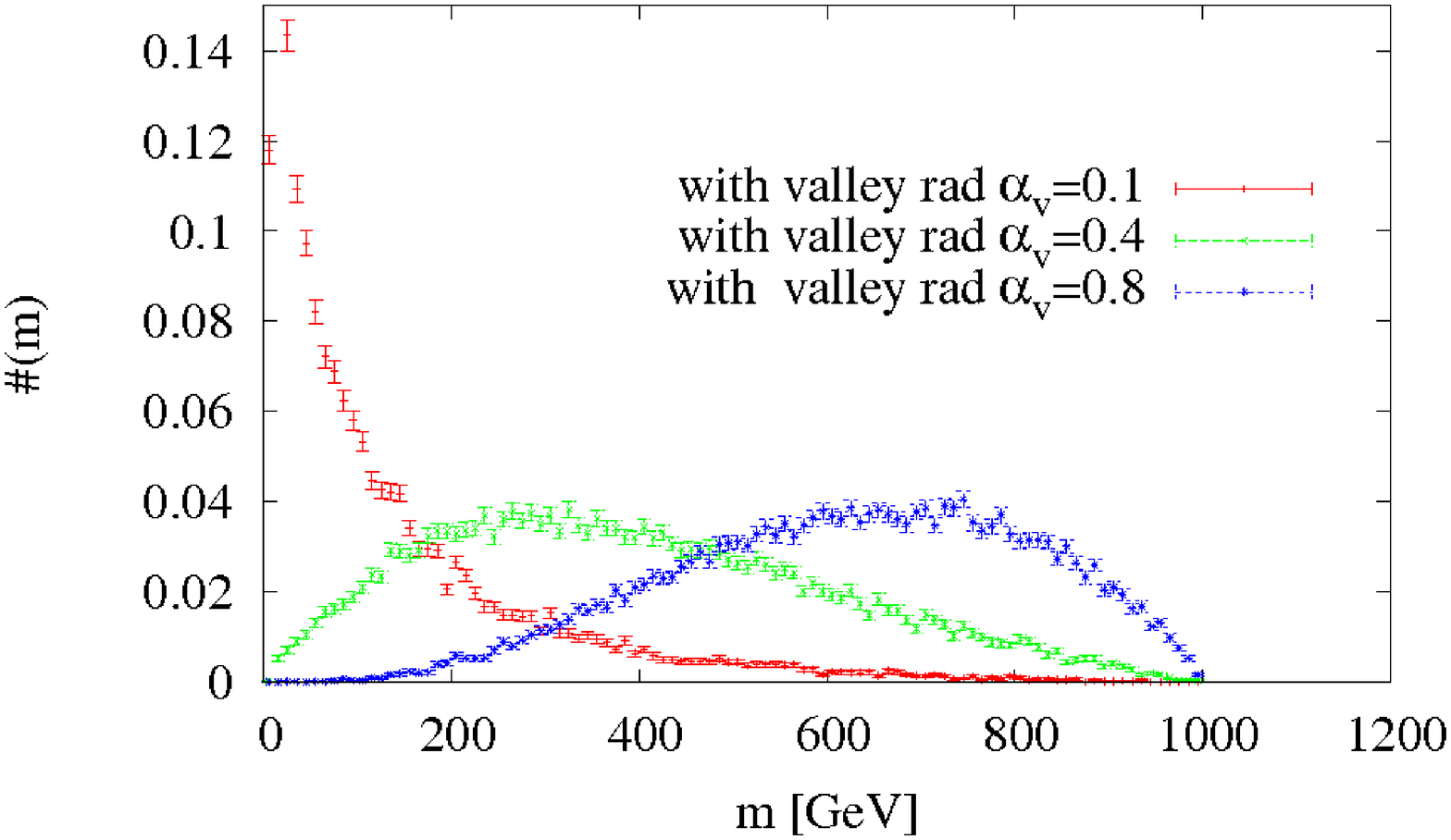, width=.315\textwidth, angle=270}
\caption{Left: the effective invariant mass distribution for the $q_v$ at LHC for $\alpha_v=~0.1,0.4,0.8$. $M_{D_v}=300$~GeV, nominal $q_v$ mass $M_{q_v}=10$~GeV and  $\sqrt{s} = 7$~TeV. Right: the effective invariant mass distribution for the $q_v$ at LHC for  $\alpha_v=~0.1,0.4,0.8$. $M_{D_v}=1$ TeV, nominal $q_v$ mass  $M_{q_v}=10$~GeV and $\sqrt{s} = 14$~TeV. Notice how the mean value of the distribution shifts from the bare mass $m_{q_v}=10$ GeV towards $m^{\max}_{q_v}=m_{D_v}$ GeV as the coupling constant grows.}\label{fig:inv_mass}
}

The hidden radiation affects the visible particle kinematics through two
mechanisms. The first one, interleaved radiation from the $E_v$ or $D_v$, we
already discussed. The second is radiation off the $q_v$ after $E_v/D_v$
decay. This causes the invariant mass for the system made out of $q_v$ and the 
radiated $g_v$s to be larger than the on-shell $q_v$ mass. This invariant mass can be 
viewed as the off-shell mass with which the $q_v$ is produced at the decay 
vertex. In Fig.~\ref{fig:inv_mass} one may see the effects
of the valley radiation on the off-shell mass, for the $D_v \rightarrow d+q_v$
case. As the hidden valley coupling $\alpha_v$ increases, the  $q_v$  radiates
more and more into the hidden sector, and the mean value of the  distribution
shifts from $M_{q_v}$ towards $M_{D_v}$. At the same time more and more energy 
is subtracted from the  visible recoiling particle, in this case the $d$ and
its system of emitted gluons.

The effect is more obvious when the mass difference  $M_{D_v}-M_{q_v}$ is
large, since more phase space is available for the radiation. 
 
The size of the deviations induced  by these two combined mechanisms is very much dependent on the collider, as we already stressed above. In the next two sections we will discuss the various  cases separately. 

\section{Effects of $SU(3)_c$ radiation at $e^+ e^-$ colliders}
\label{sec:CLIC}

We begin by studying the $e^+ e^- \rightarrow \gamma^*/Z \rightarrow \bar{E}_v E_v $, scenario, which allows for many simplifications compared to the quark case, and therefore offers a convenient warmup. For the ILC with $\sqrt{s}=800$ GeV and  an assumed integrated luminosity of $L=200$ fb$^{-1}$ per year an $M_{E_v}=300$ GeV translates into about 80000 $\bar{E}_v E_v $ pairs.

\subsection{Collisions in the center-of-mass frame}

To illustrate the principles, as a very first step we will neglect bremsstrahlung and beamstrahlung.  We then only need to consider two types of interactions, electromagnetic and valley $SU(3)_v$ radiation in the final state, with coupling constants $\alpha$ and $\alpha_v$. No fragmentation or hadronization need to be taken into account. 

Since the center of mass (CM) of the collision is at rest, there is a clean relationship between the mass of the hidden valley $q_v$,  $M_{q_v}$, and that of the communicator $M_{E_v} $. In the absence of radiation (hidden or standard), this can be  inferred from the distribution of the energy of the emitted electrons, in particular from the upper endpoint of this distribution, describing the electron maximum energy. This is obtained when the electron is emitted in the same direction as the $E_v$ is moving in, with the $q_v$ in the opposite direction. One may use this maximization condition to derive the relationship between $M_{q_v}$ and $M_{E_v} $.

In the rest frame of the $E_v$, neglecting the electron mass,
\ba
&P_{E_v}&=(M_{E_v},0,0,0)~,\nonumber\\
&P_{q_v}&=\left( \frac{M^2_{E_v}+M^2_{q_v}}{2
     M_{E_v}} ,0,0,-\frac{M^2_{E_v}-M^2_{q_v}}{2
   M^2_{E_v}}\right)~,\nonumber\\
&P_{e}&=(\frac{M^2_{E_v}-M^2_{q_v}}{2 M_{E_v}},0,0,\frac{M^2_{E_v}-M^2_{q_v}}{2 M_{E_v}})~.
\ea
Assuming the boost to the CM rest frame is at an angle $\theta$ with respect to the $e$ direction in the $E_v$ rest frame, the electron energy will be given by
\be
E^{\prime}_e = \gamma (E_{e} + \beta |\mathbf{p}_{e}| \cos\theta)
= \frac{\sqrt{s}}{4}\left(1-\frac{M^2_{q_v}}{M^2_{E_v}}\right)
\left( 1+\sqrt{1-\frac{4M^2_{E_v}}{s}}\cos\theta\right),
\ee
where $\cos\theta = \pm 1$ gives the upper and lower edge of the energy spectrum.
If the decay is assumed isotropic, $\mathrm{d}\mathcal{P}/\mathrm{d}\cos\theta =$
constant, the electron energy spectrum is flat between the limits.

So if one can measure the maximum and minimum energy $E^{\prime}_e$, one may solve for $M_{E_v}$ and $M_{q_v}$. Fig. \ref{fig:ejets} shows the energy distribution $E_e$ of the electrons produced  with and without hidden radiation.
In the latter case the spectrum is shifted to lower values, as the hidden sector takes a bigger
fraction of the available energy, by radiation off both the $E_v$ and the $q_v$. The endpoints remain the same, as there is 
always a fraction of events where radiation is negligible. As we have assumed
a modest width of 1 GeV for the $E_v$ there is a tiny tail beyond the expected edge. (We could cope with a wide range of widths, but have picked values in the GeV range, so that the possibility of a Breit-Wigner-shaped mass broadening is not overlooked, while still maintaining a credible simulation in terms of resonance diagrams only.) The key point to observe, however, is how the upper ``shoulder'' is softened
by the hidden radiation. Thereby a precision measurement of this region 
would offer a direct check on the amount of hidden radiation. 
At the lower end, QED cascades such as $e^- \to e^-\gamma \to e^-e^+e^-$ contribute to the spectrum, but are easily eliminated if only the highest-energy 
lepton is considered, right side of Fig. \ref{fig:ejets}, or at least only the highest two. We should clarify that the electron energy studied in this section includes photons emitted near the electron direction, since we here include a Durham ``jet'' algorithm that clusters photons within a 3 GeV $p_T\sin \theta/2$ distance of the electron.

\DOUBLEFIGURE[t]  
{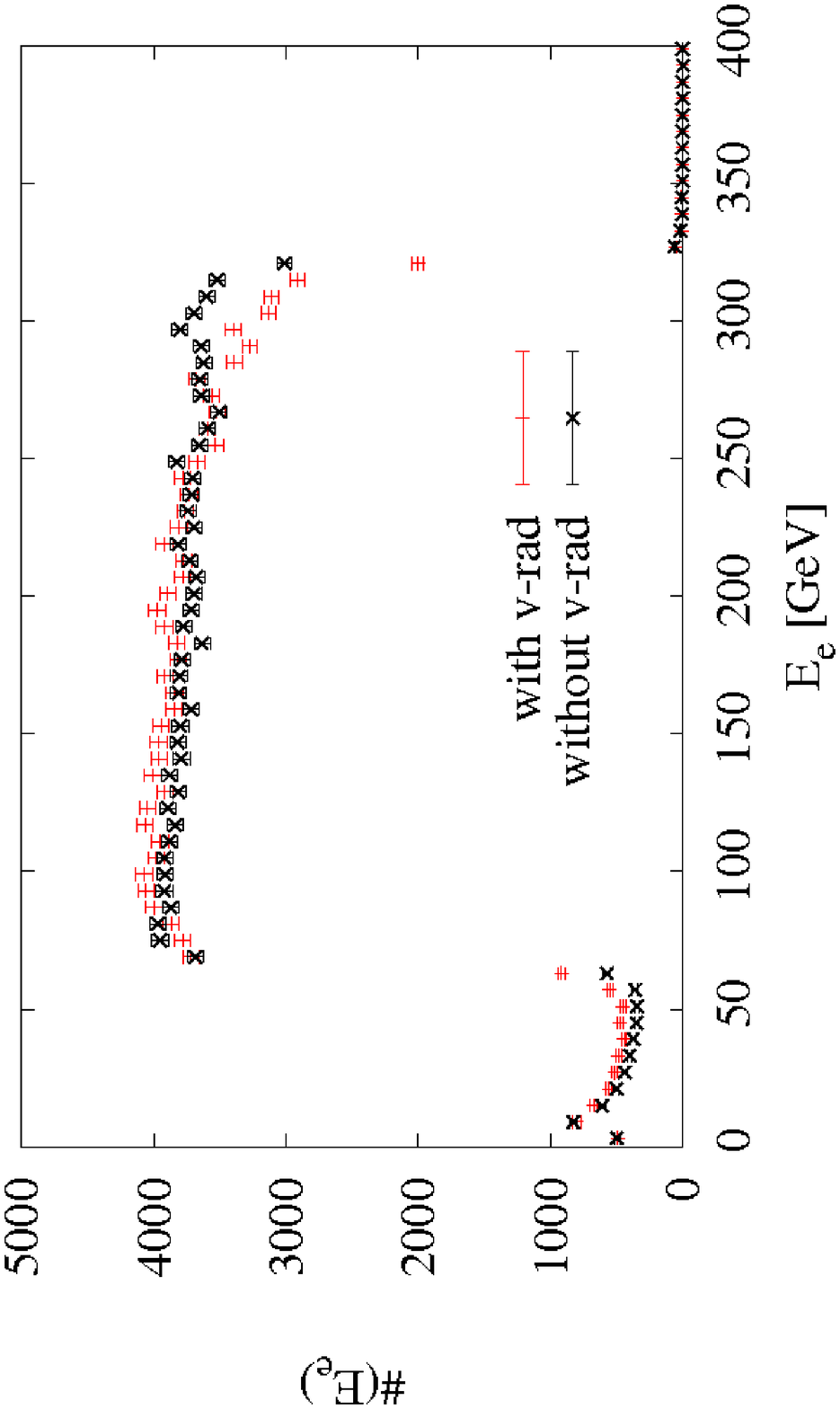,scale=0.31, angle=270}
{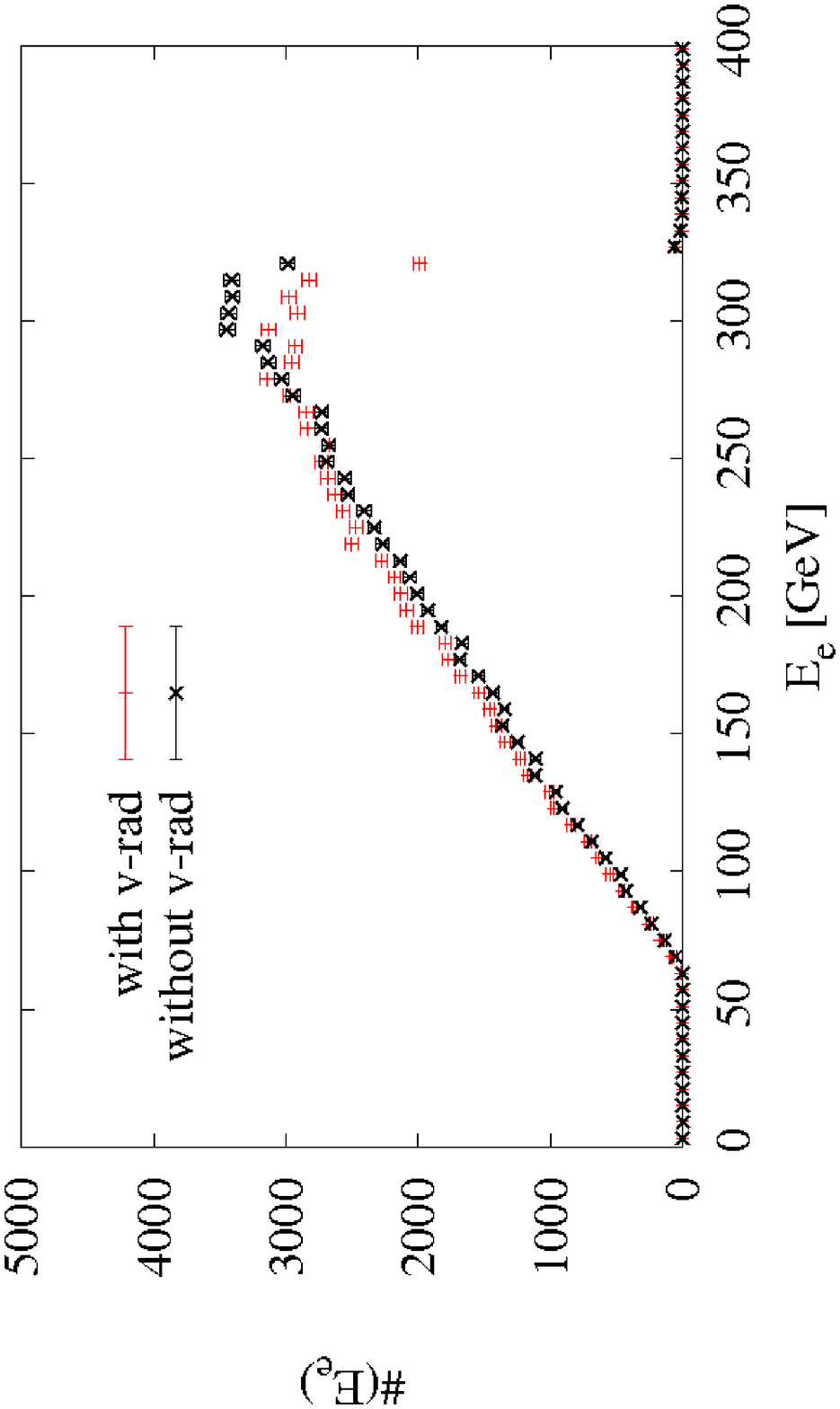,scale=0.31,angle=270}
{the energy distribution of the visible electrons. Looking at the upper shoulder of the distribution, the upper (black) and lower (red) curves give the prediction for $E_v \rightarrow  e^- q_v $  when hidden valley radiation is not or is take into account; in both cases electromagnetic radiation is included. Center of mass energy $\sqrt{s}=800$ GeV, $M_{E_v}=300$ GeV, $M_{q_v}=50$ GeV and $\alpha_v=0.05$. Number of events per 6 GeV bins, luminosity $L=200$ fb$^{-1}$. The error is purely statistical.}
{The energy distribution of the most energetic electron in each event, under the same conditions.\label{fig:ejets}}

Whether and how well one would actually be able to observe these endpoints  will be very model and detector dependant. Regardless of the background or detector sensitivity, we expect that the endpoints of the distribution will have low statistics, given that they correspond to extreme kinematical configurations. Most likely, one will need to rely on data points in the shoulder region to fit the curve and extrapolate the endpoint $E^{\prime}_{e,\max}$. These shoulder data points would be  the ones most affected by the radiation, so the mass $M_{E_v}$ and $ M_{q_v}$ inferred from them would be significantly different when hidden radiation is included. On the one hand, the curve corresponding to having valley radiation is always softer than the one without, so some mean of the threshold region
will give too low an endpoint. On the other hand, if one only tries e.g.\ a linear fit, the shape of the fall-off in the threshold region would suggest too high an endpoint. A readiness to include a parametric shape for the endpoint region, that takes into account a tuneable radiation contribution, will help ensure a better extraction of the relevant mass parameters.

Notice that the curve corresponding to having valley radiation always lies below the one without, implying that the value of $E^{\prime}_{e, \max}$ with the radiation would always be higher than the one without.  This is a reflection of the two mechanisms we mentioned in the previous section: first, the the valley gluons $g_v$  subtract energy from the $E_v$, ultimately subtracting it from the $e$s, and second, when they are emitted by the $q_v$, they change its effective mass, $M^{\mathrm{eff}}_{q_v}>M_{q_v}$, as one may see in Fig.~\ref{fig:inv_mass}, again subtracting energy from the decay $e$.

%where we stress the uniqueness of the v-radiation signal
Would it be possible to describe the curves with hidden radiation using a model without it, but with different mass parameters $M_{E_v}$ and $M_{q_v}$? Fig. \ref{fig:massvariations} shows the effects of changing the invisible particle mass $M_{q_v}$ in model with and without radiation.  The "fingerprint'' of the v-radiation is clear: a softening in the shoulder of the distribution which leaves the endpoints fixed. A simple change in the mass parameters of the model without hidden radiation (in this case $M_{q_v}$) changes the endpoints and leaves the sharp drop of the shoulder unchanged.

\FIGURE[t]{\epsfig{file=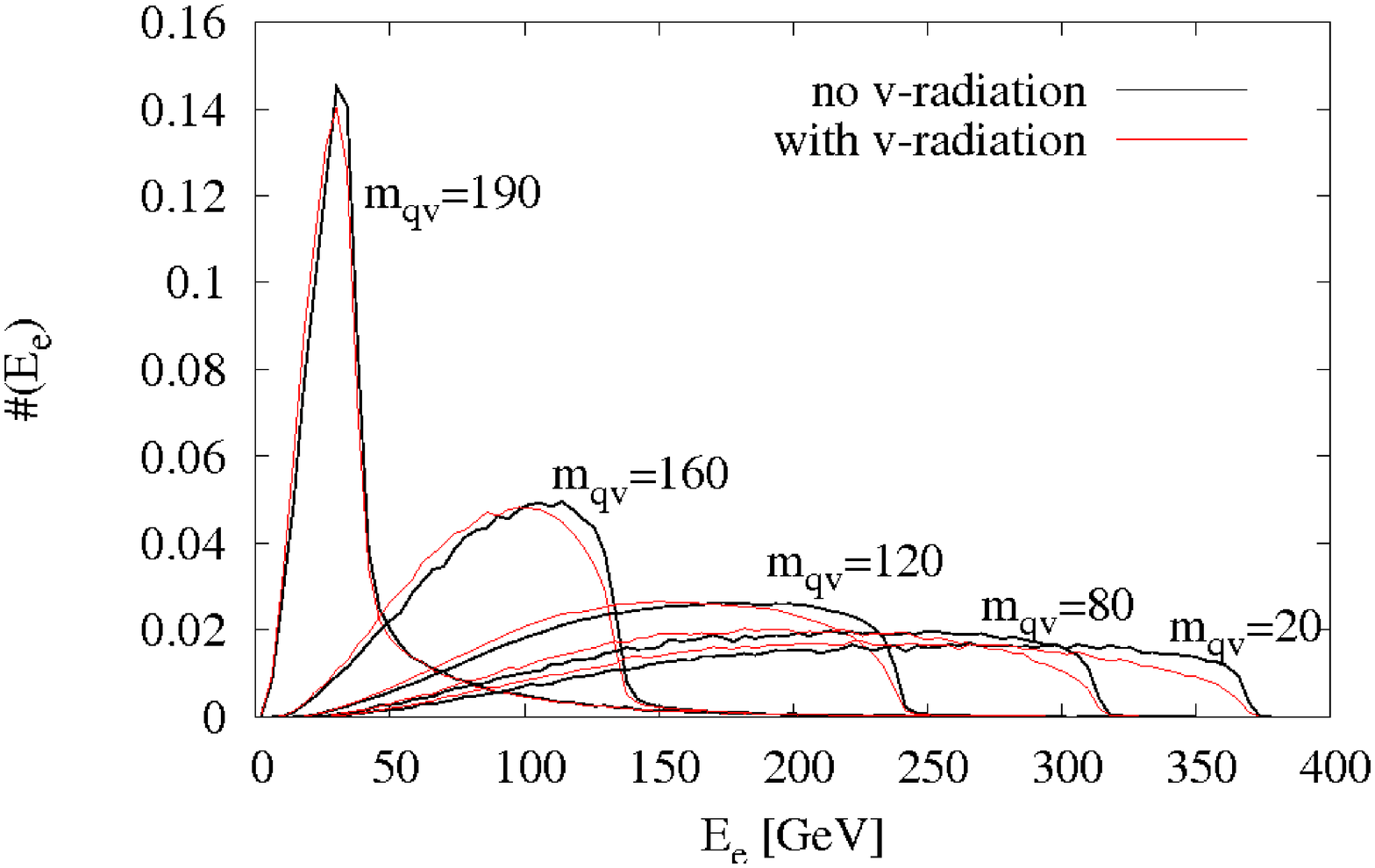,width=0.5\textwidth, angle=270}
\caption{We compare the energy distributions of the most energetic electron emitted in each event, in the case $\sqrt{s}=800$ GeV. The communicator mass is fixed at $M_{E_v}=200$ GeV, while $M_{q_v}$ is allowed to vary between 20 and 190 GeV. The valley gauge coupling is fixed at  $\alpha_v=0.1$  in order to isolate the mass dependence. Each endpoint corresponds to two curves, the lower one being the one with and the top one the one without hidden valley radiation.}
\label{fig:massvariations}
}

Notice how  so long as the mass difference $M_{E_v}-M_{q_v}>40$ GeV one may always distinguish between any two curves with and without radiation. This  of course $\alpha_v$ dependent.

\FIGURE[t]{\epsfig{file=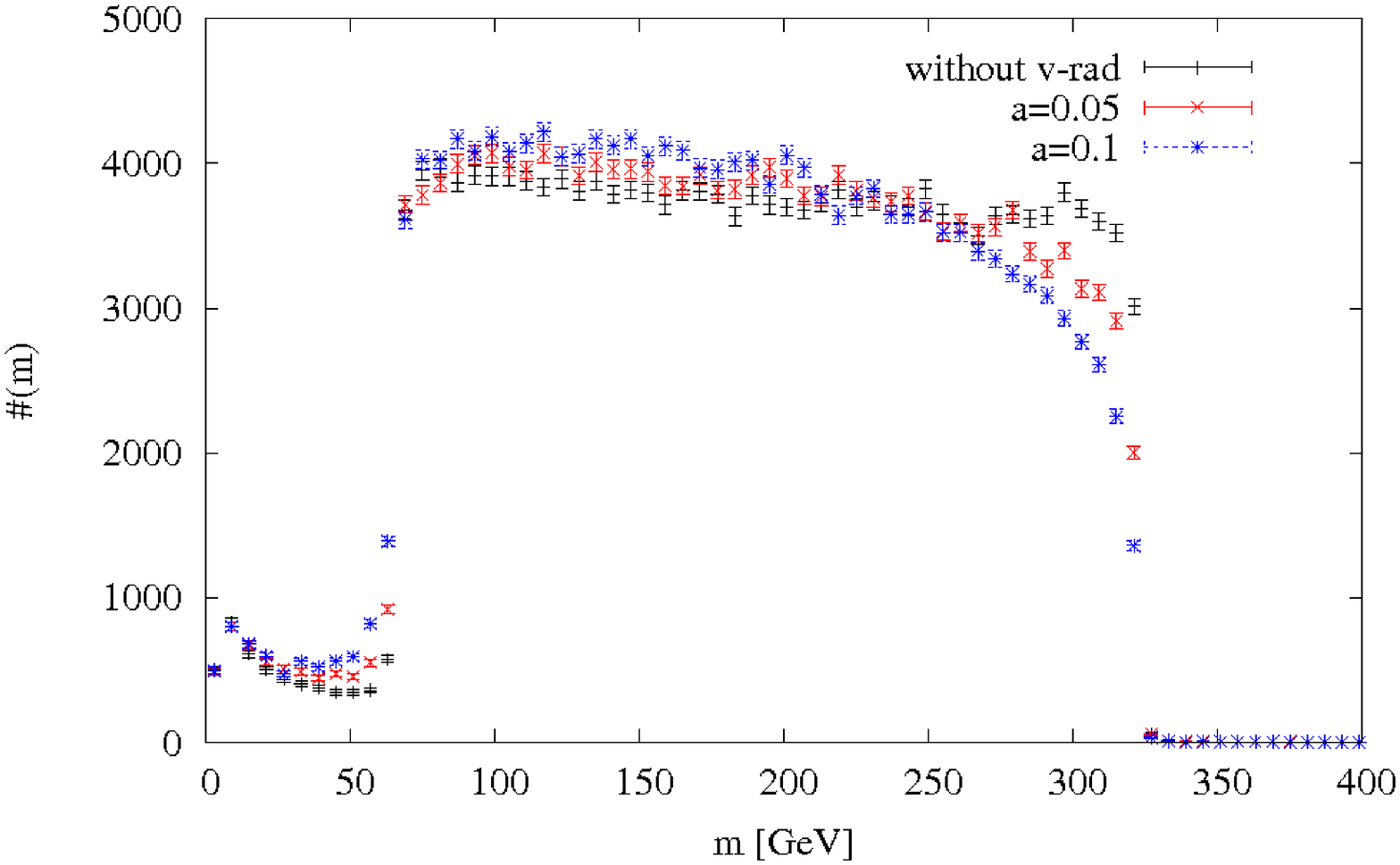,width=0.5\textwidth, angle=270}
\caption{$\alpha_v$ dependence of the energy of the visible electrons. The (black) squared-off curve corresponds to the model with no valley radiation, the uppermost if one looks at the shoulder. Below it are the curves corresponding to valley coupling constant $\alpha_v= 0.05$ and $\alpha_v= 0.1$. Number of electrons per 6 GeV bin, $\sqrt{s}=800$~GeV, $M_{E_v}=300$ GeV and $M_{q_v}=50$ GeV.}
\label{fig:alphavariations}
}

In Fig. \ref{fig:alphavariations} instead one can see the  $\alpha_v$ dependence of  the energy distribution for the visible electrons in each event.  Notice how even for a coupling as low as $\alpha_v=0.05$ the effects of the v-radiation on the shoulder region are already sizeable.

There are some parameter regions (e.g.\ when the $E_v$-to-$q_v$ mass splitting is small) where the shape of the distribution of the hardest leptons  is no longer conclusive in distinguishing between a model with and one without valley radiation.
 In this case one may consider other observables which have an ``orthogonal'' dependence on the valley parameters. We studied $\eta$, linearized sphericity $S$ and the number of emitted leptons, for example. There is no unique strategy in this case, one must perform  a case by case study of the different observables in the different parameter regions to determine which one displays the largest separation between the model with and the model without radiation. As a general rule, we found that three observables were normally sufficient to distinguish the two.

\subsection{Collisions not in the center-of-mass frame: MT2}
\label{subsec:MT2}

We now consider the effect  of initial-state radiation (ISR).  This causes 
unobservable radiation, mainly along the beamline, but also some transverse 
kicks. Beamstrahlung is highly machine-dependent and thus not included, but 
is purely longitudinal. The methods we will introduce to handle bremsstrahlung 
also automatically handle beamstrahlung with little or no degradation of 
performance, so from now on we will not address the latter specifically.

For our theoretical studies, in order to avoid the clustering of ISR $\gamma$ radiation with the leptons  coming from the hard interaction, we apply a cut on the $\eta>5$. The symmetry of the system now being cylindrical, we also changed the clustering algorithm to the cylindrical fastjet \cite{Cacciari:2005hq}.

The major consequence of ISR is that the collision now no longer happens in the CM rest frame, with the information connected to the $p_z$, the momentum along the beampipe, no longer available. In this case it is convenient to introduce a new variable called Cambridge MT2, see \cite{Lester:1999tx}.

The MT2 variable was invented precisely to treat  events in which the new particles are pair-produced  and then each decay into one particle that is directly observable and another particle whose existence may only be inferred from from missing \emph{transverse} momenta.

This observable is somewhat inspired by the transverse mass $m_T$ used at hadron colliders to measure the mass of the $W$ boson in the decay $W \rightarrow e \nu$. The neutrino escapes detection, its only trace in the detector being missing momentum.  In this case one can construct the variable
\be
m^2_T=2(E^e_T \hspace{2pt}\not \hspace{-4pt}E^N_T-\mathbf{p}^{e}_T\cdot \not \hspace{-3pt}\mathbf{p}^N_T).
\ee
Here $E_T$ is defined as $E_T=\sqrt{m^2+\mathbf{p}^2_T}$, although in this particular case the electron and neutrino masses can be neglected, of course. The $m^2_T$ variable has the property that
\be
m^2_T\le m^2_W
\ee

If there is enough statistics to ensure that the kinematic configuration corresponding to the maximum is hit, this gives a measurement of (a lower bound on)  the $W$ mass.
Analogously, one may build a variable called MT2, with the property that its
upper bound describes the mass of the communicators, i.e.\ the particles that
were pair-produced.

\DOUBLEFIGURE[t]
{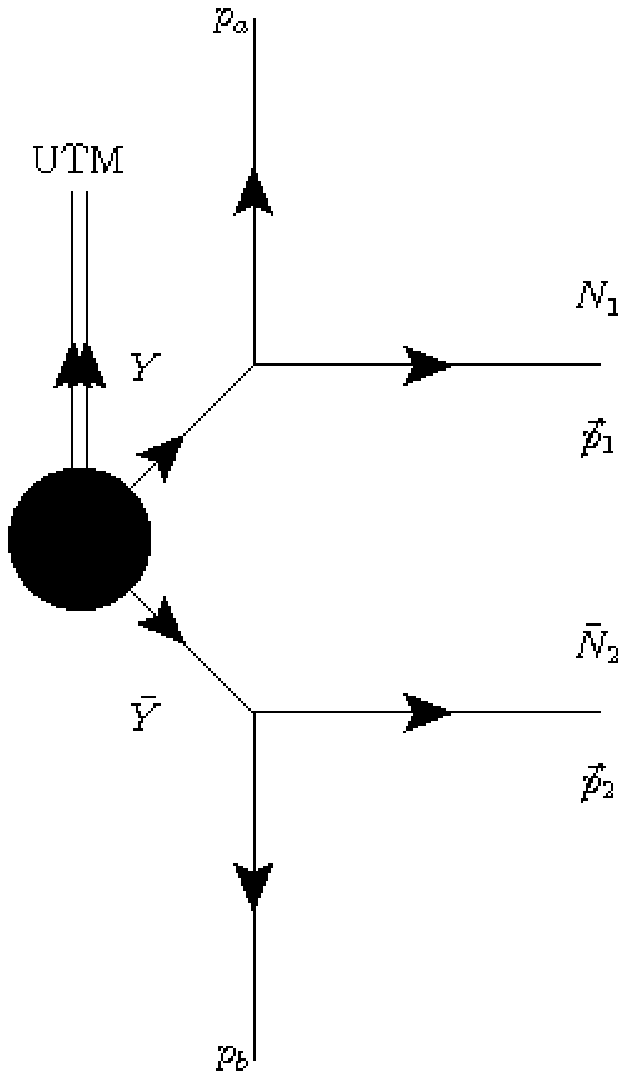,width=0.3\textwidth}
{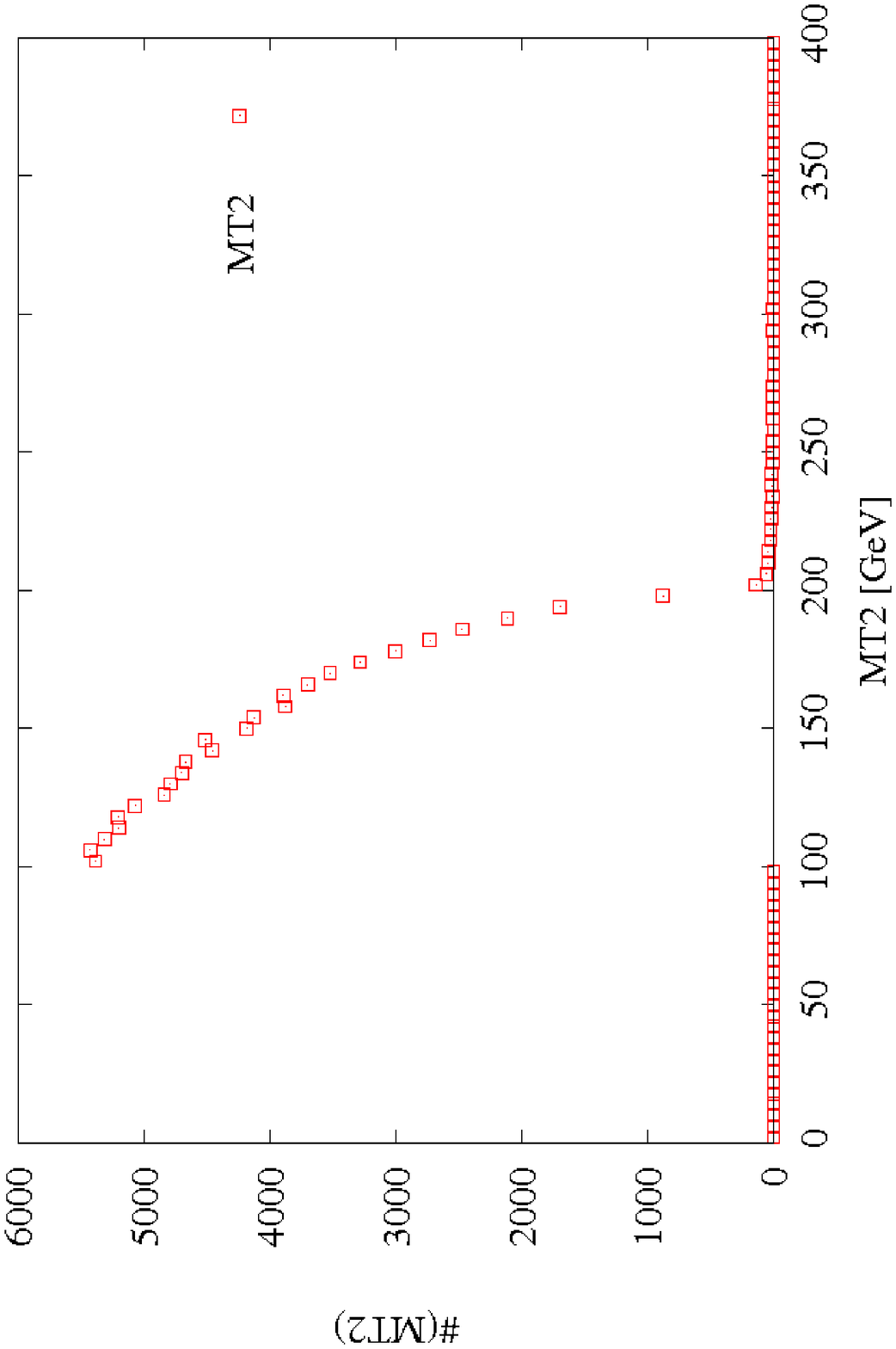,width=0.4\textwidth, angle=270}
{The MT2 diagram for the shortest, simplest decay chain . Two communicator particles are pair-produced and each decay into a visible ($a$ or $b$) and an invisible ($N_1$ and $N_2$) particle. Upstream transverse momentum (before the decay) is also possible, but it must be known. Longitudinal momentum information is not available.\label{fig:MT2_dgrm}}
{An example of the MT2 method applied to the communicator mass $M_{E_v}$. Histogram of the MT2 values obtained for a communicator mass $M_{E_v}=200$ GeV and  $M_{q_v}$ and an invisible particle mass $M_{q_v}=100$ GeV. The small tail in the distribution is due to the $\Gamma=1$ GeV spread in the mass distribution (more about his in the following).\label{fig:typicalMT2}}

Now consider the process described in Fig. \ref{fig:MT2_dgrm}. Two particles $Y$ are pair-produced, then each of them decays into a visible particle ($a$ or $b$ in the figure) and one that escapes detection, called $N_{1,2}$.
 One cannot use the transverse momentum in this case since there are two particles escaping detection, both contributing to the missing transverse momentum $\not \hspace{-3pt} \mathbf{p}_T$.  The observable MT2 is defined as
\be
\mbox{MT2} \equiv \min_{ \not \mathbf{p}_{T1}+ \not \mathbf{p}_{T2}= \not  \mathbf{p}_T}  \left[ {\max} \left\{ m^2_T(\mathbf{p}_{Ta}, \not \hspace{-3pt} \mathbf{p}_{T1}),m^2_T(\mathbf{p}_{Tb},\not \hspace{-3pt} \mathbf{p}_{T2}) \right\} \right].
\ee
where $\not \hspace{-3pt} \mathbf{p}_{T1,T2}$ are all the possible 2-momenta taken away by the $N$s, such that their sum gives the observed missing momenta, $\not \hspace{-3pt} \mathbf{p}_{T1}+\not \hspace{-3pt} \mathbf{p}_{T2}= \not \hspace{-3pt} \mathbf{p}_T$.

MT2 coincides with the mass of the communicator $Y$, i.e. MT2 has a maximum, when for both communicator decays the visible and the invisible particles are produced at the same rapidity and 
\be
\left(
\frac{\mathbf{p_a}}{E_a}-\frac{\mathbf{p_1}}{E_1}\right)\propto \left(\frac{\mathbf{p_b}}{E_b}-\frac{\mathbf{p_2}}{E_2}
\right) ~.
\label{eq:MT2_edge}
\ee

For the studies in this article we used a particularly simple version of the MT2 algorithm \cite{Cheng:2008hk}, the source code of which can be downloaded from\\
\underline{http://daneel.phyics.uc.davis.edu/~Cheng:2008hk/mt2-1.01a/test}.\\ 
A more sophisticated algorithm is described  in \cite{lester_web}. 
The simpler method is based on the use of "kinematic constraints'' \cite{Bachacou:1999zb,Allanach:2000kt,Hinchliffe:1996iu,Cheng:2008hk}
\ba
\mathbf{p}_1^2  = \mathbf{p}_2^2&=&\mu^2_N\nonumber\\
(\mathbf{p }_1+\mathbf{p}_a)^2  &=& (\mathbf{p}_2+\mathbf{p}_b)^2=\mu_Y^2\nonumber\\
p^x_1+p^x_2 &=& \not\hspace{-2pt}{p}^x\nonumber\\
p^y_1+p^y_2 &=& \not\hspace{-2pt}{p}^y
\ea 

In the case of two invisible and at least two visible particles as  in Fig. \ref{fig:MT2_dgrm} the two methods actually coincide \cite{Cheng:2008hk}. 

The inputs of the MT2 method are $m_N$, $m_a$, $m_b$, $\mathbf{p}^a_T$, $ \mathbf{p}^b_T$ and $\not\hspace{-3pt}\mathbf{p}_T$. Notice that $m_a$ and $m_b$ may change quite substantially from event to event, since they each correspond to the invariant masses of the clustered visible particles (in this case the lepton and the photons) of each branch. The output of the MT2 method is one single MT2 value per event. Fig.~\ref{fig:typicalMT2} illustrates a typical use of the MT2 variable. If one histograms the MT2 values over a large number of events, the upper edge of this distribution gives a lower limit on the communicator mass $M_Y$. 

Whether the event rate in the upper-edge kinematic region defined in eq.~(\ref{eq:MT2_edge}) is large enough to be able to extract the endpoint MT2$^{\max}=M_Y$, for a given luminosity, depends of course on the interactions. Even more than for the energy variable in the previous section, it is not unlikely that MT2$^{\max}$ might have to be extrapolated from points in the shoulder region.

%where we explain why we picked MT2 rather than another observable
There are many other methods to determine mass relations between the the new particles, \cite{Hinchliffe:1996iu,Tovey:2000wk}, just to cite some. Some of these are very closely related to MT2, such as \cite{Cho:2007qv}. Some of these require cascade decay chains, or make assumptions about the new particles involved in the decay chain being on shell, or require high luminosity. Where this information is actually available, one should of course make use of it, \cite{Cheng:2007xv,Cheng:2008mg}.

The assumptions in this study, though, are that each of the identical decay chains consists of a single two-body decay and that the integrated luminosity, at least for the LHC at 7~TeV study, might be rather low (1 fb$^{-1}$). These effectively preclude  the use of many of the above methods.  

\FIGURE[t]{
\epsfig{file=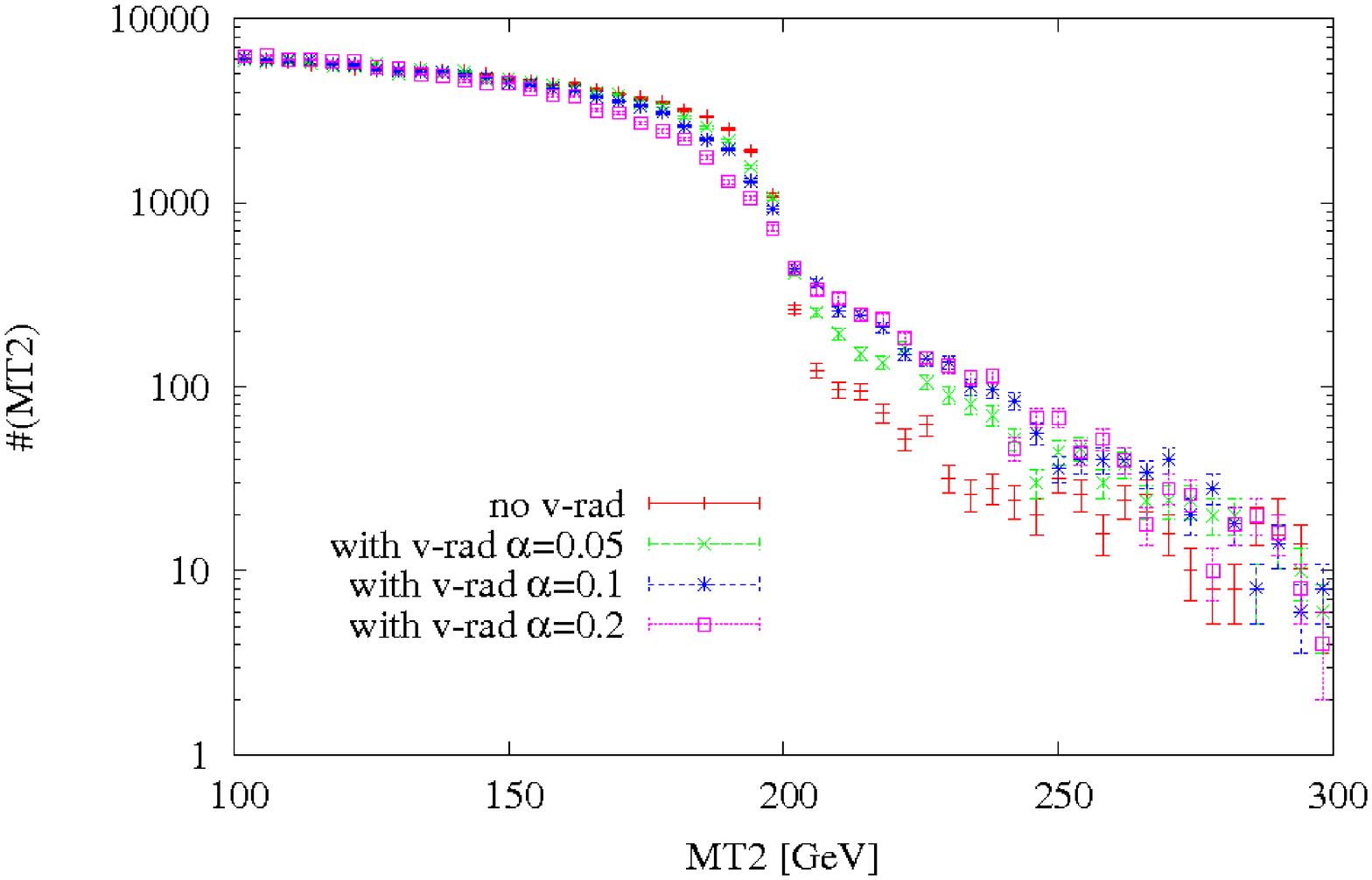,scale=0.31,angle=270}
\epsfig{file=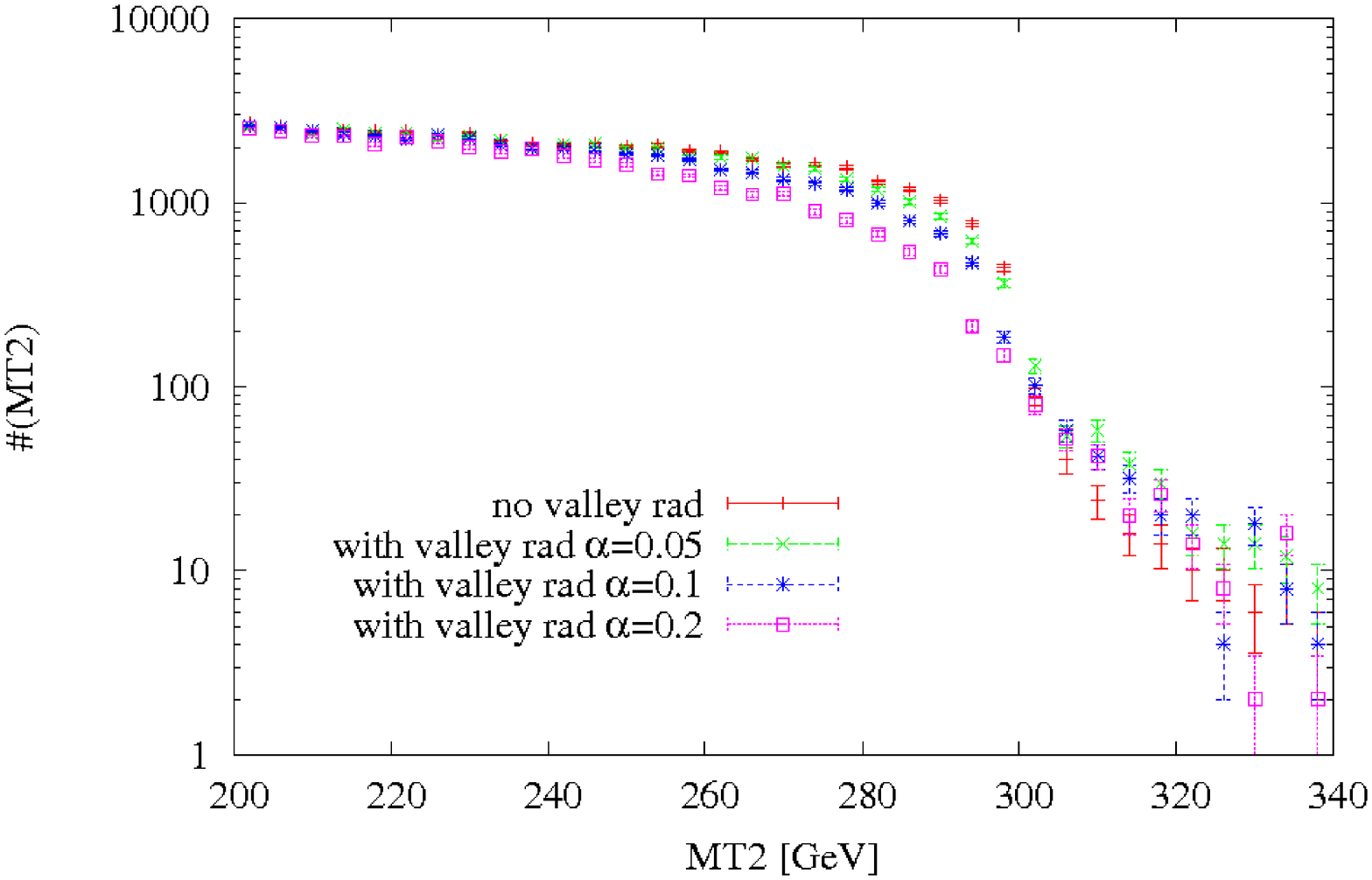,scale=0.31,angle=270}
\caption{The effect of valley radiation on the variable MT2 for different  $\alpha_v=0.05, 0.1, 0.2$ values (close-up on the shoulder region) at CLIC. The distributions were obtained assuming a luminosity $L=1$ fb$^{-1}$, $\sqrt{s}=1$ TeV and $M_{q_v}=50$ GeV. Left: the effect for a hypothetical $M_{E_v}=200$ GeV. Right: the effect of valley radiation on MT2 for a hypothetical  $M_{E_v}=300$ GeV.}
\label{fig:MT2_a}
}

\FIGURE[ht]{
\epsfig{file=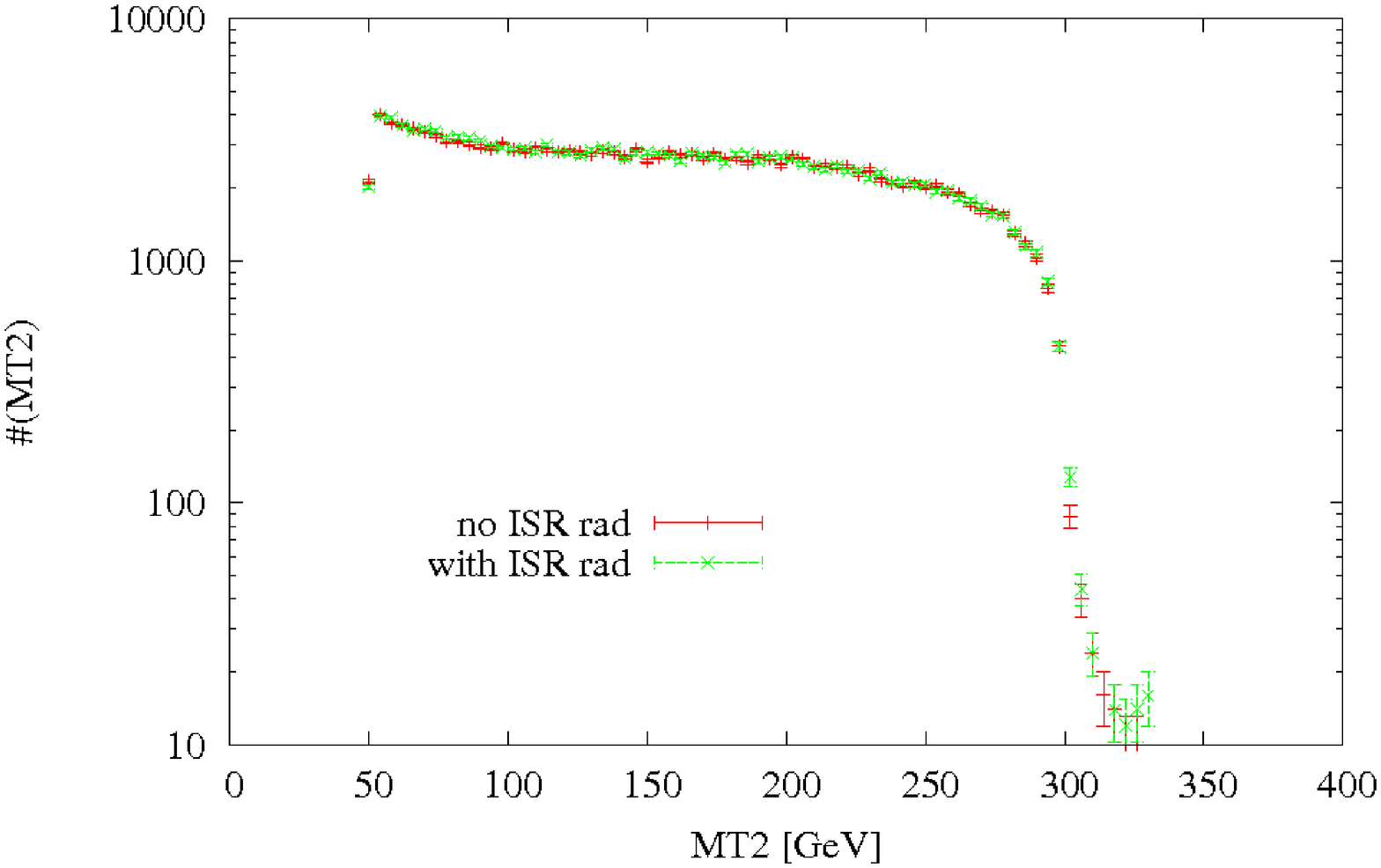,scale=0.32, angle=270}
\caption{The effects of initial-state radiation on the MT2 distribution and on the value of the $M_{E_v}$ mass one infers. Number of events per 4 GeV bin, for an $e^+ e^-$ collider with $\sqrt{s}=1$ TeV, $L=1$ fb$^{-1}$, $M_{E_v}=300$ GeV, $\Gamma_{E_v}=2$ GeV, $M_{q_v}=50$ GeV. All other effects have been switched off.}
\label{fig:ISR}
}

In Fig. \ref{fig:MT2_a} one may see the effect of the valley radiation on the MT2 distribution for different $\alpha_v$ values and communicator mass parameters. The interesting region is again represented by the "shoulder'' of the distribution. Notice how the amount of invisible radiation, and thus the effect on MT2$^{\max}$,  increases with $M_{E_v}-M_{q_v}$, analogously to what happens in the energy distributions. The size of these effects may  be compared with the effects coming from  ISR, in Fig. \ref{fig:ISR}.

\FIGURE[t]{
\begin{minipage}[b]{0.4\linewidth}
\hspace{-30pt}
\epsfig{file=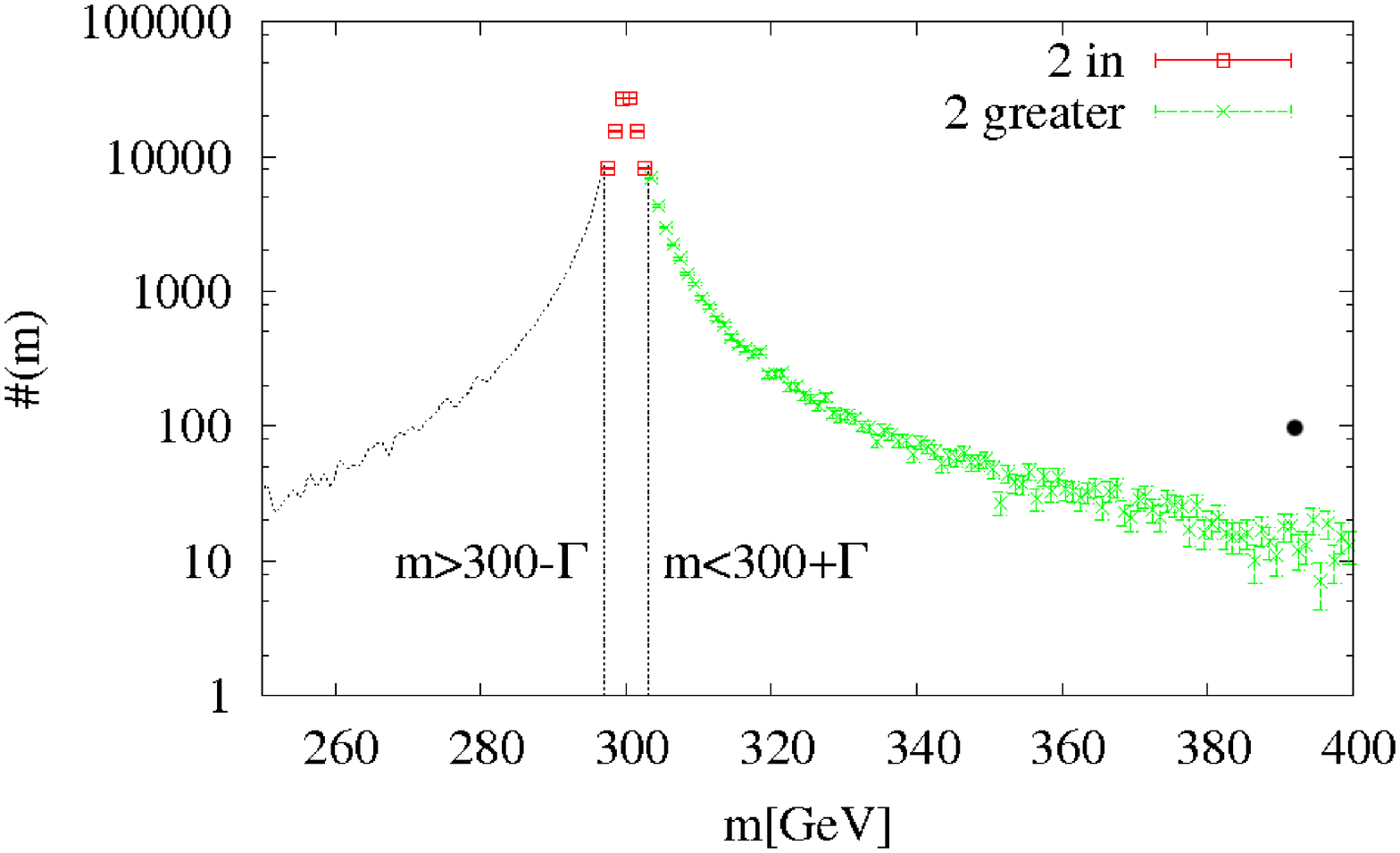,scale=0.315, angle=270}
\end{minipage}
\hspace{0.5cm}
\begin{minipage}[b]{0.4\linewidth}
\epsfig{file=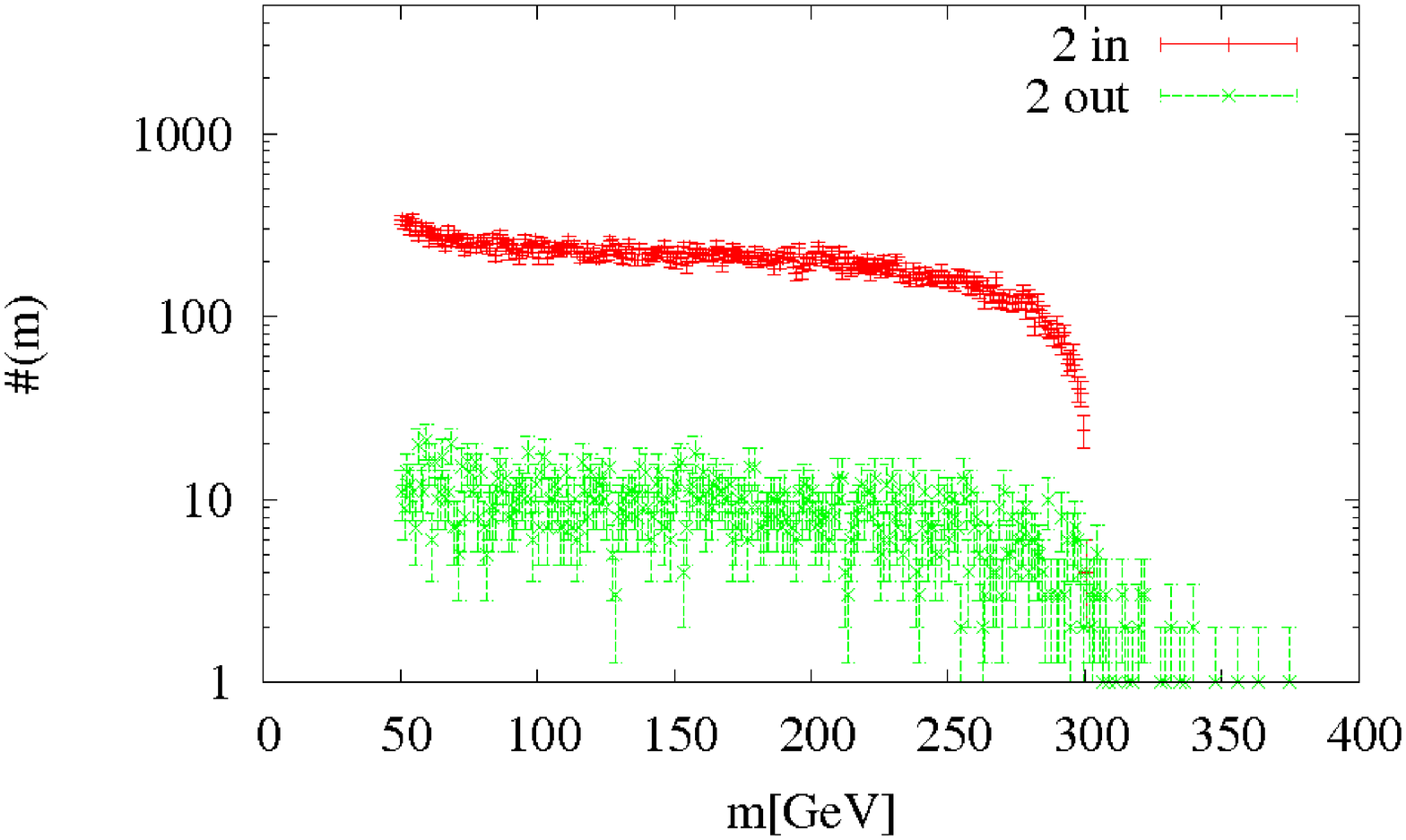,scale=0.315, angle=270}
\end{minipage}
\caption{Left: the (right side of the) Breit-Wigner mass distribution of the communicator  $E_v$ with an $M_0=M_{E_v}=300$ GeV and a width $\Gamma_{E_v}=3$ GeV. Right: MT2 distribution for the same points.  The upper (red) points correspond to considering only $E_v$s having masses  within the $\vert m_{E_v}-M_{E_v}\vert < \Gamma_{E_v}$ GeV interval. The lower points (green) describe the  MT2 points corresponding to both  $E_v$s having a mass greater than $M_{E_v}+ \Gamma_{E_v}$. All other effects have been switched off. Number of events per 1 GeV bin, $\sqrt{s}=800$ GeV, $M_{q_v}=50$ GeV, integrated luminosity $L=200$ fb$^{-1}$. The error is purely statistical.}
\label{fig:gamma}
}
 
The MT2 distribution might present a tail, due to the Breit-Wigner spread $\Gamma$ that we allow for, see Fig. \ref{fig:gamma}. As one may see, all the MT2 data points which lie above $\mbox{MT2}^{{\max}}=M_{E_v}=300$ GeV correspond to $E_v$s which actually have a larger mass than the nominal $M_{E_v}$. 

%where we discuss how to calculate m_n
As we already stated above, $m_N$, the mass of the invisible particle, is an input parameter.
MT2$^{\max}$ only gives one single relation between $M_Y$ and $m_N$. 
Depending upon the decay chain topology and the presence or not of upstream transverse momentum (UTM, in the following, may come from ISR or from previous decays), 
there are different strategies to determine both masses simultaneously: the MT2 ``kink'' method  \cite{Cho:2007qv}, the invariant mass endpoint \cite{Gjelsten:2004ki,Costanzo:2009mq,Burns:2009zi,Matchev:2009iw} or 
the constrained kinematic method  \cite{Cheng:2007xv}, the polynomial intersection method \cite{Cheng:2008mg}, and $\mathbf{p}_T$ reconstruction  \cite{Kawagoe:2004rz,Nojiri:2007pq,Webber:2009vm}, just to cite some possibilities. Most methods, however \cite{Gjelsten:2004ki,Costanzo:2009mq,Burns:2009zi,Matchev:2009iw, Cheng:2007xv},  \cite{Cheng:2008mg}, require longer decay chains (at least two two-body decays) or a special topology, such as 4 on-shell intermediate resonances  \cite{Cheng:2007xv} or 5 or more on-shell intermediate resonances \cite{Cheng:2008mg}.

If each decay chain consists of a single two-body decay, where the visible one may or may not be a composite of visible particles, as described in Fig. \ref{fig:MT2_dgrm},  one may use the MT2 "kink method'' \cite{Cho:2007qv} to fix the value of $m_N$, i.e. exploit the fact that MT2$^{\max}$ as a function of the invisible particle trial mass $\mu_N$ has a "kink'' for $\mu_N=m_N$. The authors of \cite{Barr:2010zj} point out that in order to have a substantial change in the gradient $\frac{d MT2^{\max}}{d \mu_N}\vert_{\nu_N=m_N}$, there must be substantial event-by-event changes, though. This can be triggered by substantial $O(M_Y)$ differences in the $\nu_N$, caused by the visible system being a collection of two or more particles, or by a large UTM. Otherwise the kinematics is so constrained that the gradients for $\mu_N< m_N$ and $\mu_N> m_N$ have to be the same, and no kink is possible.

If a sizable UTM $\mathbf{p_T}$ is present, one may use the MT2$\perp$ method \cite{Konar:2009wn}. This method uses the fact that $N(\mu_N)$, the number of times the $MT2(\\mu_N,\mathbf{p}_T)$ is larger than $MT2(\mu_N,0)$, has a minimum for $\mu_N=m_N$. The advantage of using this method rather that MT2kink is that $MT2(\mu_N,0)$ may be calculated analytically and measured using the \emph{whole} data sample, regardless the $\mathbf{p_T}$. This may be shown by using the fact that $MT2(\mu_N,0)$ corresponds to  MT2$^{\max}\perp(\mu_N)$, 
\ba
\mbox{MT2}^{\max}_{T\perp}=\min_{\not P_\perp=p_{1T\perp}+p_{2T\perp}} \left[ \max \left \{\mbox{M}^2_{1T\perp},\mbox{M}^2_{2T\perp}\right \} \right],
\ea
the one-dimensional analogue of MT2, where
\ba
\mbox{M}^2_{iT\parallel}&=&m^2_i+\mu^2_N+2(E_{i T\parallel}E^N_{iT\parallel}-p_{iT\parallel}\cdot p^N_{iT\parallel})\nonumber\\
        M^2_{iT\perp}&=&m^2_i+\mu^2_N+2(E_{iT\perp}E^N_{iT\perp}-p_{iT\perp}\cdot p^N_{iT\perp})\nonumber,
\ea
and where $\perp$ and $\parallel$  refer to the projections of the $\mathbf{p_T}$ along the direction of the UTM. 

We will not discuss further the different methods to extract the two new particle masses, but refer the interested reader to the proceedings from the TeV 2009 \cite{Brooijmans:2010tn} conference and to the review \cite{Barr:2010zj}. We however wish to make a few remarks about the impact that valley radiation might have on these observables. Consider the MT2$\perp$ case, for example.

%where we discuss the impact of the v-radiation on the MT2_t
In the presence of valley radiation one needs to consider two sources of deviations. Firstly,  the tails coming from the interleaved radiation mechanism, see the MT2 distributions in Fig. \ref{fig:MT2_a}. Secondly, as discussed in the previous section, in the presence of valley radiation, we expect the mean value of the invisible particle $q_v$ invariant mass to shift from its Breit-Wigner central value $\langle M^{\mathrm{eff}}_{q_v} \rangle =\mu_N$ towards the communicator mass $M_{E_v}$ (or $M_{D_v}$) value. We will show in subsection \ref{subsec:LHC_14} that the MT2 distribution one obtains may be significantly affected, see Fig. \ref{fig:inv_mass} for th LHC case. The $\mbox{MT2}^{\max}_{T\perp}(\mu_N)$ should be similarly affected. The number of events having $\tilde{M}_Y(\mu_N,\mathbf{P}_T)>\tilde{M}_Y(\mu_N,0)$ would then change accordingly, as would the minimum point $\mu_N=M_N$.

We will return on the issue of the  the trial mass and the radiation in subsection \ref{subsec:LHC_14}. In the following, unless otherwise specified, the analysis will always assume $\mu_N=m_N$.

\section{Effects of $SU(3)_v$ radiation at LHC}
\label{sec:LHC}

At LHC the $D_v$ communicators are (mostly) pair-produced by $gg$  or $q \bar{q}$ fusion and decay flavour diagonally into a SM $d$ quark and a valley $q_v$. For our study we assume the $D_v$s to be spin 1/2 particles, and the $q_v$s to be scalars. As earlier this choice affects the production cross section, but now both $s$- and $t$-channel exchange are involved, which complicates the pattern.
Each $D_v$  radiates both SM $g$s and valley $g_v$s. These in turn may radiate further $g$s and $g_v$s, respectively. Once the $D_v$ has decayed, the $q$ radiates gluons, while the $q_v$ radiates $g_v$s. The amount of hidden radiation emitted depends upon the valley coupling constant $\alpha_v$ and on the mass ratio $M_{q_v}/M_{D_v}$, see Fig. \ref{fig:inv_mass}. At the LHC the communicator mass reach will be larger than for the ILC, so typically there will be more phase space available for the radiation. Both the valley gluons radiated by the $D_v$ and those radiated by the $q_v$ have an impact on the visible particle distributions. The lighter the particle, the lower the cut-off scale for the radiation however, so it will be the $q_v$ that radiates the most, as before.

%list of complications and relative weight of effects
When compared with the CLIC case, the LHC scenario presents  several complications. Firstly one needs to convolute the production cross section with parton distribution functions. Thus the hard interaction --- the production of the $D_v$ 
pair --- no longer happens in or close to the center-of-mass rest frame. 
In this case it is crucial to consider longitudinally boost invariant 
observables such as MT2. Secondly, both initial- and final-state QCD 
radiation are more intense than the QED one is for the ILC case, resulting 
in a considerably larger upstream transverse momentum and an increased 
misassignment of radiation. Thirdly, there is an underlying-event activity 
that gives rise both to a generic low-$p_{\perp}$ background and to occasional 
further hard partons that may be confused with the ones related to the valley
process. Fourthly, the partons hadronize into more-or-less well-defined 
hadronic jets, the reconstruction of which introduces further smearing 
of the relevant kinematic distributions. And finally, the set of possible 
background processes is much more varied and challenging to suppress. In our study we will take into account the first four points, but leave
the last one to the experimental community, where already a large number of
background-suppression techniques have been developed for various scenarios.

%should comment on the presence of the decay coupling size in the following

\subsection{LHC with 7 TeV}
The LHC will initially be running at of $\sqrt{s}=7$ TeV and it is expected to deliver 1 fb$^{-1}$ of data in 2010--2011. Under these conditions we need to consider much lower masses  $M_{D_v}$ for the communicator than for the ultimate energy and luminosity case, in order to have large enough production cross sections, see Table \ref{tab:sigmas}.

Based on the above discussions, we choose to study the  MT2 distribution, and specifically its dependence on the v-radiation and on the $\alpha_v$ value. In Fig. \ref{fig:MT2_adep_LHC_7} we have plotted the $\alpha_v$ dependence for three different mass values $M_{D_v}=300, 400$ and 500 GeV.  The larger the mass difference $M_{D_v}-M_{q_v}$ the more phase space is available for the radiation. The smaller the $M_{q_v}$, the lower the cut-off on the momenta, so the larger the amount of soft radiation. Given the low statistics, the (purely statistical) error bars on the endpoints are rather large, and even in the shoulder of the distribution it is hard to distinguish the curve with an  $\alpha_v=0.1$ from the curve with no radiation for the 300 GeV mass.  In the intermediate case $M_{D_v}=400$ GeV, we need to have a rather strong $\alpha_v=0.2$ coupling before the two curves can be separated.
 
\FIGURE[ht]{
\begin{minipage}[b]{\linewidth}
\centering
\epsfig{file=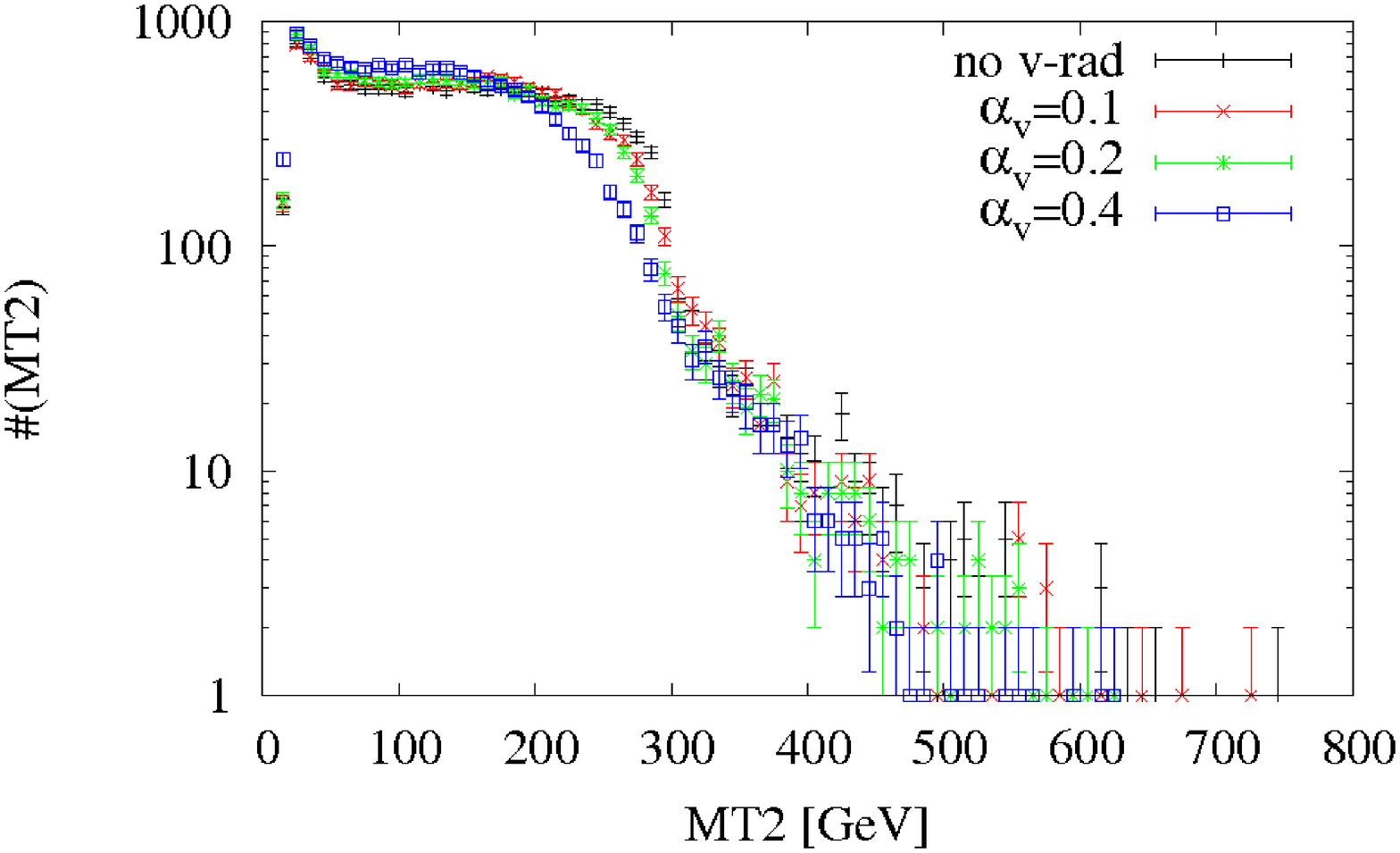,width=0.36\textwidth,angle=270}\\
%\caption{default}
%\label{fig:figure1}
\vspace{0.5cm}
\end{minipage}
\begin{minipage}[b]{\linewidth}
\centering
\epsfig{file=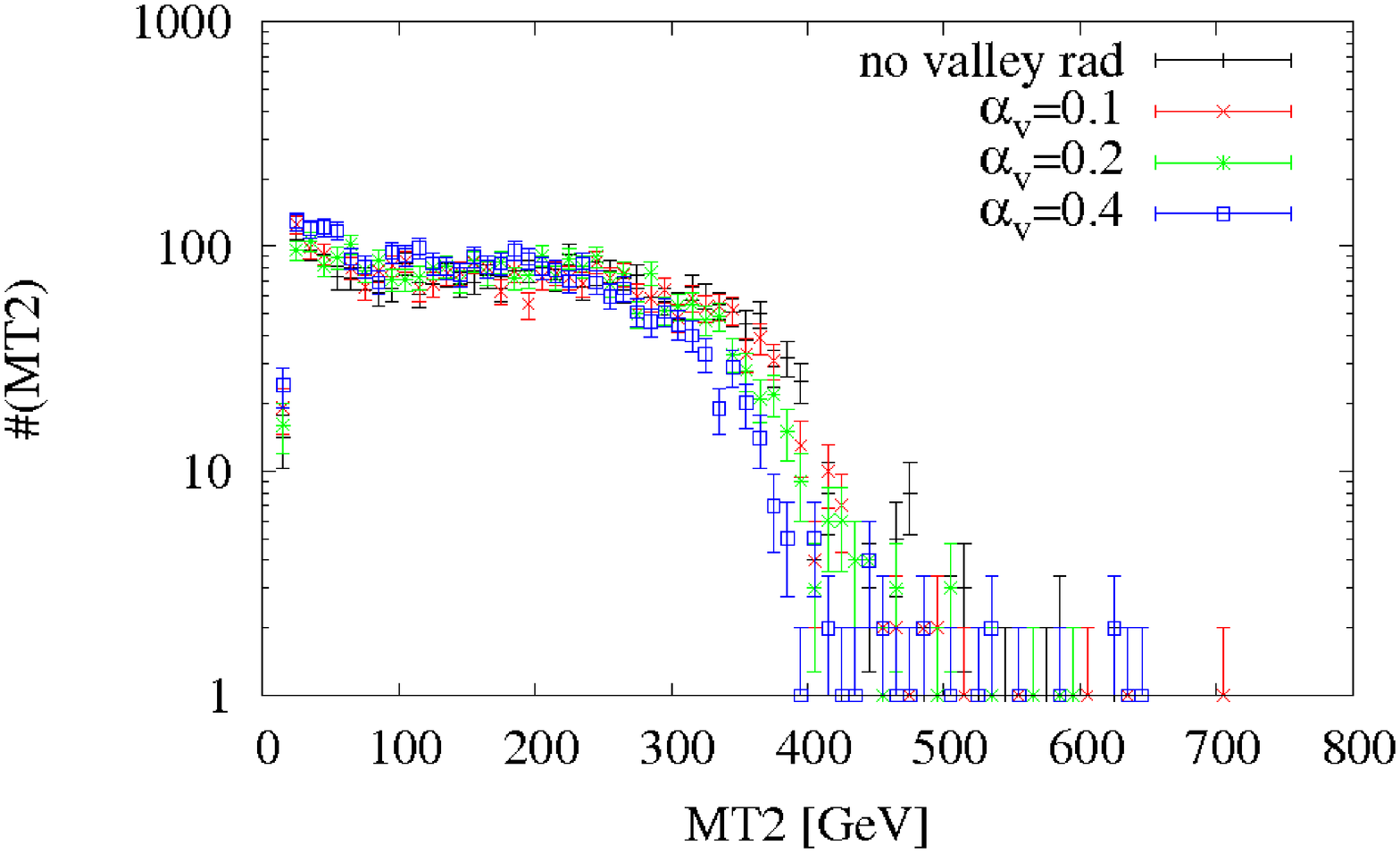,width=0.36\textwidth,angle=270}\\
%\caption{default}
%\label{fig:figure2}
\vspace{0.3cm}
\end{minipage}
\begin{minipage}[b]{\linewidth}
\centering
\epsfig{file=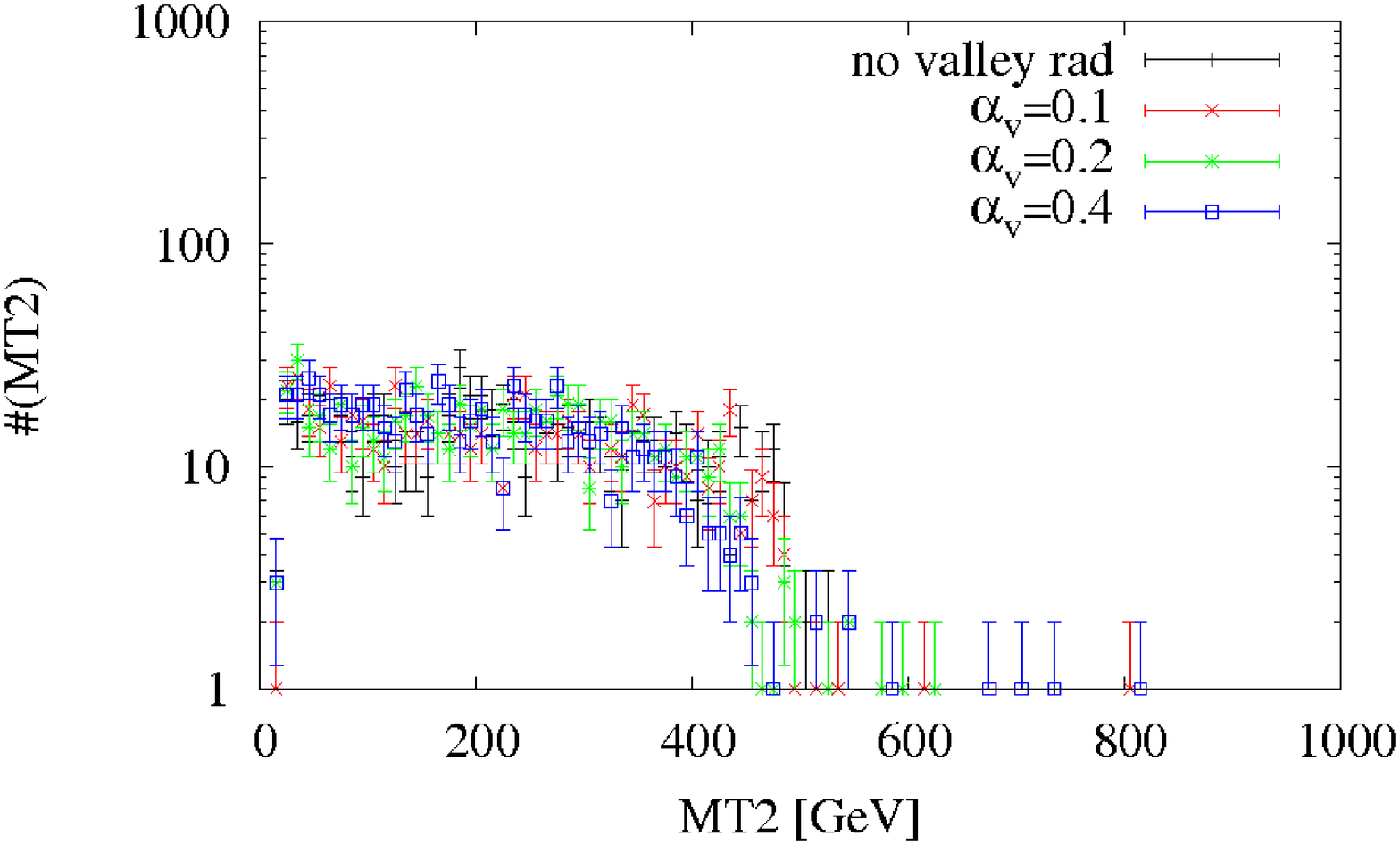,width=0.36\textwidth,angle=270}
\caption{The dependence of the MT2 distribution on the $\alpha_v$ value for $M_{D_v}=300, 400,500$ GeV and $M_{q_v}=10$ GeV. The black curve corresponds to having no valley radiation. The red, green and blue curve correspond to $\alpha_v=0.1, 0.2, 0.4$ respectively. For the first 2 years LHC is assumed to run $\sqrt{s}=7$ TeV and to yield an integrated luminosity of 1 fb$^{-1}$. The $y$ axis corresponds to the number of events per 10 GeV mass bin, for this integrated luminosity of 1 fb$^{-1}$.}
\label{fig:MT2_adep_LHC_7}
\end{minipage}
}

%where we discuss the MT2 dependence on the q_v mass 
In Fig. \ref{fig:MT2_mqvdep_LHC_7} we show the MT2 dependence on the  invisible $q_v$ mass $M_{q_v}$, where the trial mass $\mu_{q_v}$ is assumed to coincide with $M_{q_v}$. Whether we look at the curves with a valley radiation (valley coupling $\alpha_v=0.1$, left plot in Fig. \ref{fig:MT2_mqvdep_LHC_7}) or at the curves without hidden radiation (right plot in Fig. \ref{fig:MT2_mqvdep_LHC_7}), the data points corresponding to $M_{q_v}=10, 50$ GeV are hardly distinguishable. The independence of the MT2 on the $M_{q_v}$  value appears to be a characteristic which the radiation leaves unchanged.

\FIGURE[h]{
\hspace{-0.8cm}
\begin{minipage}[b]{0.41\linewidth}
\centering
\epsfig{file=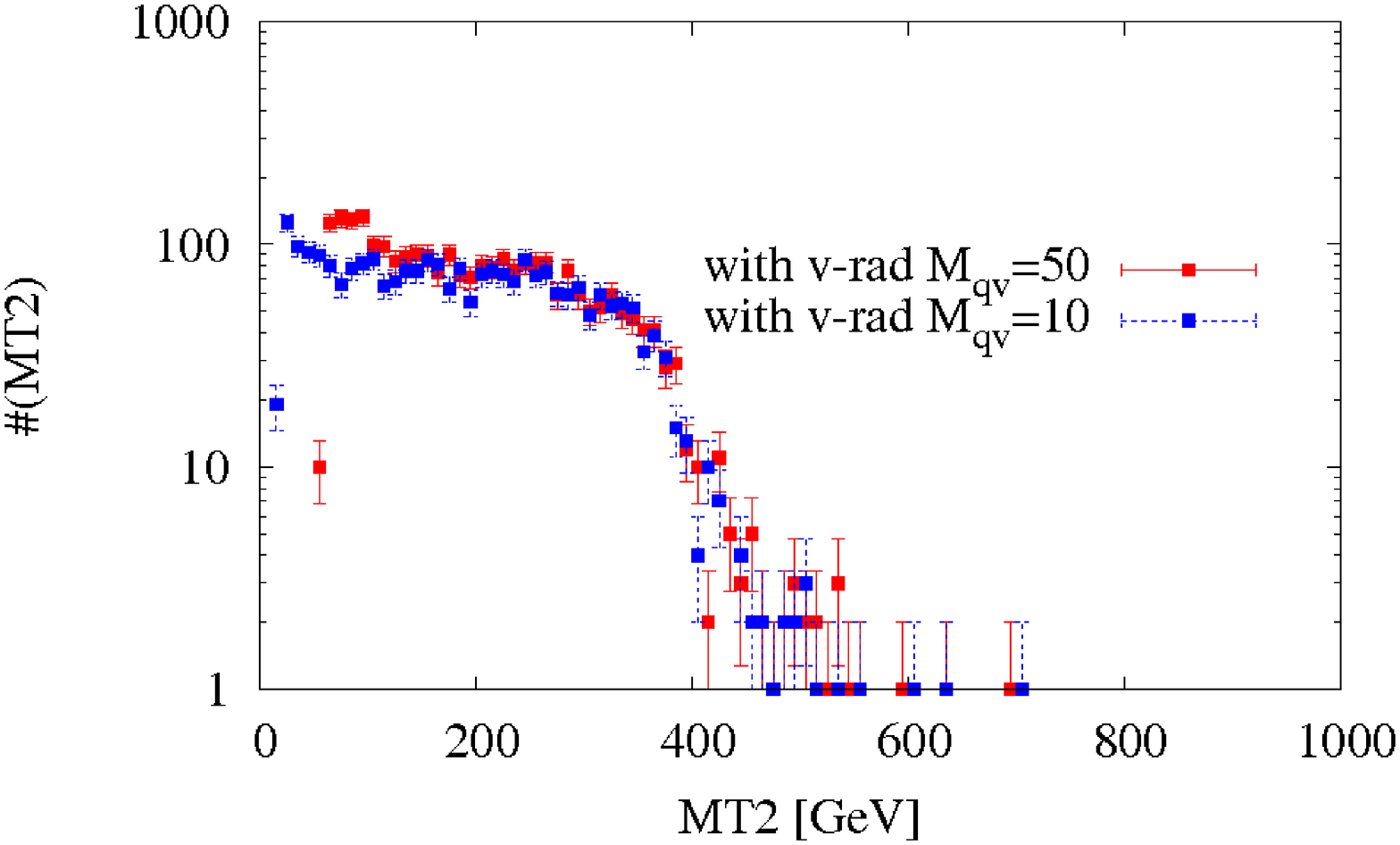,width=0.8\textwidth, angle=270}
\end{minipage}
\hspace{0.6cm}
\begin{minipage}[b]{0.41\linewidth}
\centering
\epsfig{file=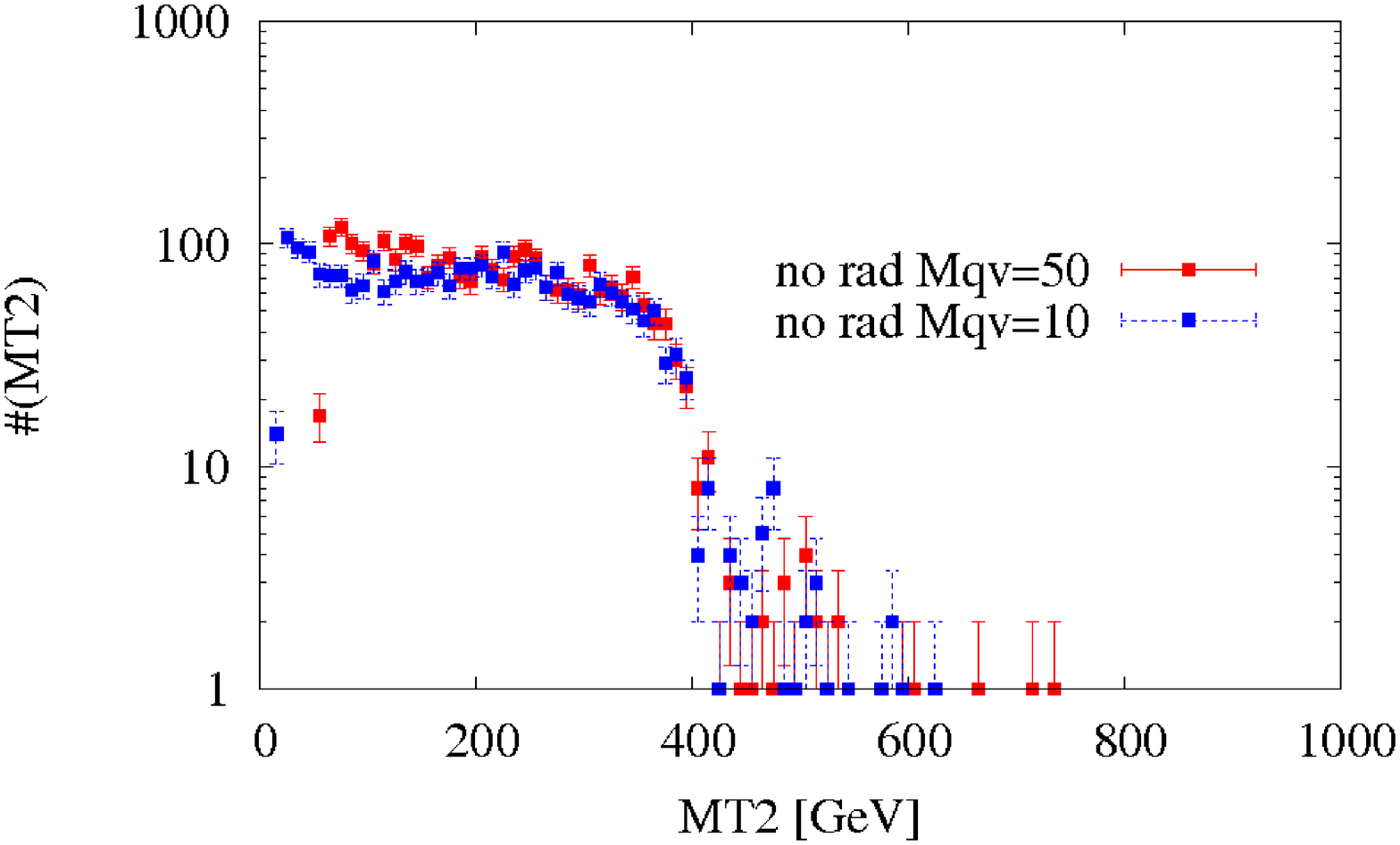,width=0.8\textwidth, angle=270}
\end{minipage}
\caption{Left: the $M_{q_v}$ dependence of the MT2 when the valley radiation has  $\alpha_v=0.1$. The blue curve corresponds to $M_{q_v}=10$ GeV and the red one to $M_{q_v}=50$ GeV. Right: The same $M_{q_v}$ dependence of the MT2 in the case of no valley radiation. $M_{D_v}=400$ GeV, $\sqrt{s}=7$ TeV and $L=1$ fb$^-1$.}
\label{fig:MT2_mqvdep_LHC_7}
}

%where we explain the independence  of the MT2 from mqv
One could argue that the fact that MT2 is hardly dependent on the $M_{q_v}$ is due to the mass ratio  $M_{D_v}/M_{q_v}$ being fairly large compared to the difference between the two $M_{q_v}$ values we considered. Would this argument still hold true when the v-radiation is larger?

%where we introduce the (wrong) concept of mqv as invariant mass
  
%$\alpha_v\le 0.2$ (we are still discussing the intermediate case $M_{D_v}=400$ GeV).
 For larger values of the $\alpha_v$ one should consider the fact that the MT2 input parameter $M_{q_v}$ corresponds to the  mass of the  $q_v$ as seen by the visible particles. The visible particle momenta, i.e. the momenta of the $d$  and the $g$ radiated by it, which enter MT2, actually correspond to  the invariant mass $M^{\mathrm{eff}}_{q_v}$ rather than the  nominal mass value $M_{q_v}$.

%where we discuss the impact of the change in the invariant mass
In Fig. \ref{fig:inv_mass} we show this invariant mass distribution  and how this changes as a function of the coupling constant $\alpha_v$.  As $\alpha_v$ grows $q_v$ emits more and more valley gluons $g_v$. The invariant mass of the $q_v+g_v$s system then grows, i.e the mean effective mass of the $q_v$ as seen from the SM $q$ shifts from the bare $M_{q_v}=10$ GeV towards $M_{q_v}^{\max}=M_{D_v}$. The energy which is left for the $d$ quark then gets smaller as the $\langle M_{q_v}^{\mathrm{eff}} \rangle \rightarrow M_{D_v}$ as shown by   
\be 
E_d=\frac{M^2_{D_v}-M^2_{q_v}}{2M_{D_v}},
\ee       
\noindent where for simplicity we have put the $m_d=0$. 

\FIGURE[ht]
{
\epsfig{file=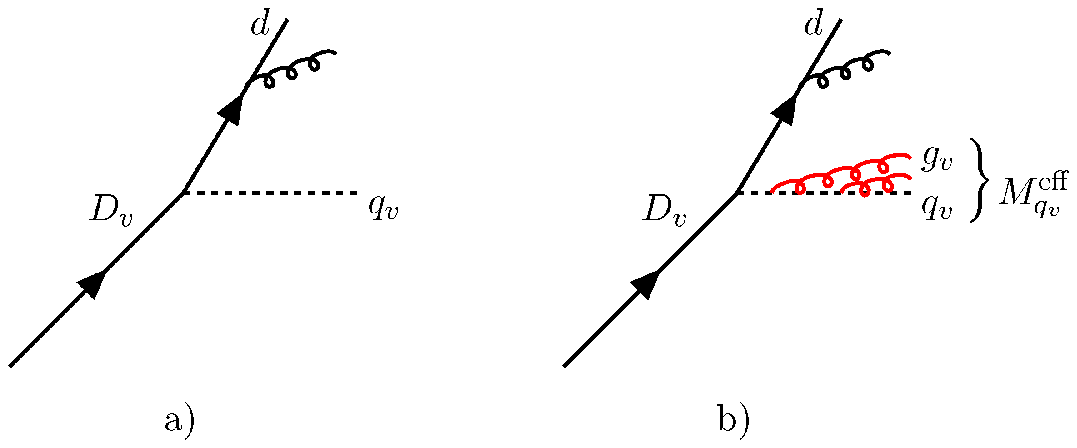,width=0.6\textwidth}
\label{fig:models}
\caption{Left: model a) decay  $D_v\rightarrow d q_v$ with no v-radiation, $M_{q_v}$ has a fixed value; right: model b) decay  $D_v\rightarrow d q_v$ with v-radiation, the system $q_v+g_v$s has an invariant mass distribution.}
}

Were this distribution a very narrow peak around $\langle M^{a,\mathrm{eff}}_{q_v}\rangle$, there would be no difference between the two cases in Fig. \ref{fig:models} so long as $\langle M^{a,\mathrm{eff}}_{q_v}\rangle=M^b_{q_v}$, i.e. between a fixed $M^b_{q_v}$ value and an invariant mass distribution. One could speculate that so long as the mean value of the invariant mass $\langle M^{\mathrm{eff}}_{q_v} \rangle \ll M_{D_v}$ the MT2 would basically remain unaffected. 

Imagine though having a  v-radiation large enough to shift the $\langle M^{\mathrm{eff}}_{q_v} \rangle$ substantially, $M^{\mathrm{eff}}_{q_v} \rightarrow M_{D_v}$, e.g. $\alpha=0.4$. One would na\"{i}vely think that replacing the $M_{q_v}$ with $\langle M^{\mathrm{eff}}_{q_v} \rangle$  would give a better description of the case with radiation. However Fig.  \ref{fig:inv_mass} shows that for $\alpha_v>0.1$ the invariant mass distribution would also have a large spread around this central value. We will show in the next section that this spread  causes further complications. It is precisely the spread in the distribution which constitutes the difference between case a) and case b), and it is this spread which is ultimately responsible for the different behaviour of the MT2 in the two cases.

\subsection{LHC with 14 TeV}
\label{subsec:LHC_14}

If we now assume that LHC will collect $100$  fb$^{-1}$ of data at center of mass energy $\sqrt{s}=$14 TeV, then one may consider larger communicator masses, $O(1\mbox{TeV})$, and still deal with a sufficient  number of events, see Table \ref{tab:sigmas}.  If the $q_v$ remains light, the mass ratio $M_{D_v}/M_{q_v}$ can be considerable and consequently the phase space available for radiation can be large. We expect the effects of the radiation to be significant. 

In Fig. \ref{fig:MT2_LHC_14} we show the dependence of the  MT2 distribution on the hidden valley coupling constant $\alpha_v$, assuming we know  the mass of the invisible particle so $\mu_{q_v}=M_{q_v}$. In this case we are describing the strong coupling regime $\alpha_v=0.1, 0.4, 0.8$.

\FIGURE[ht]{\epsfig{file=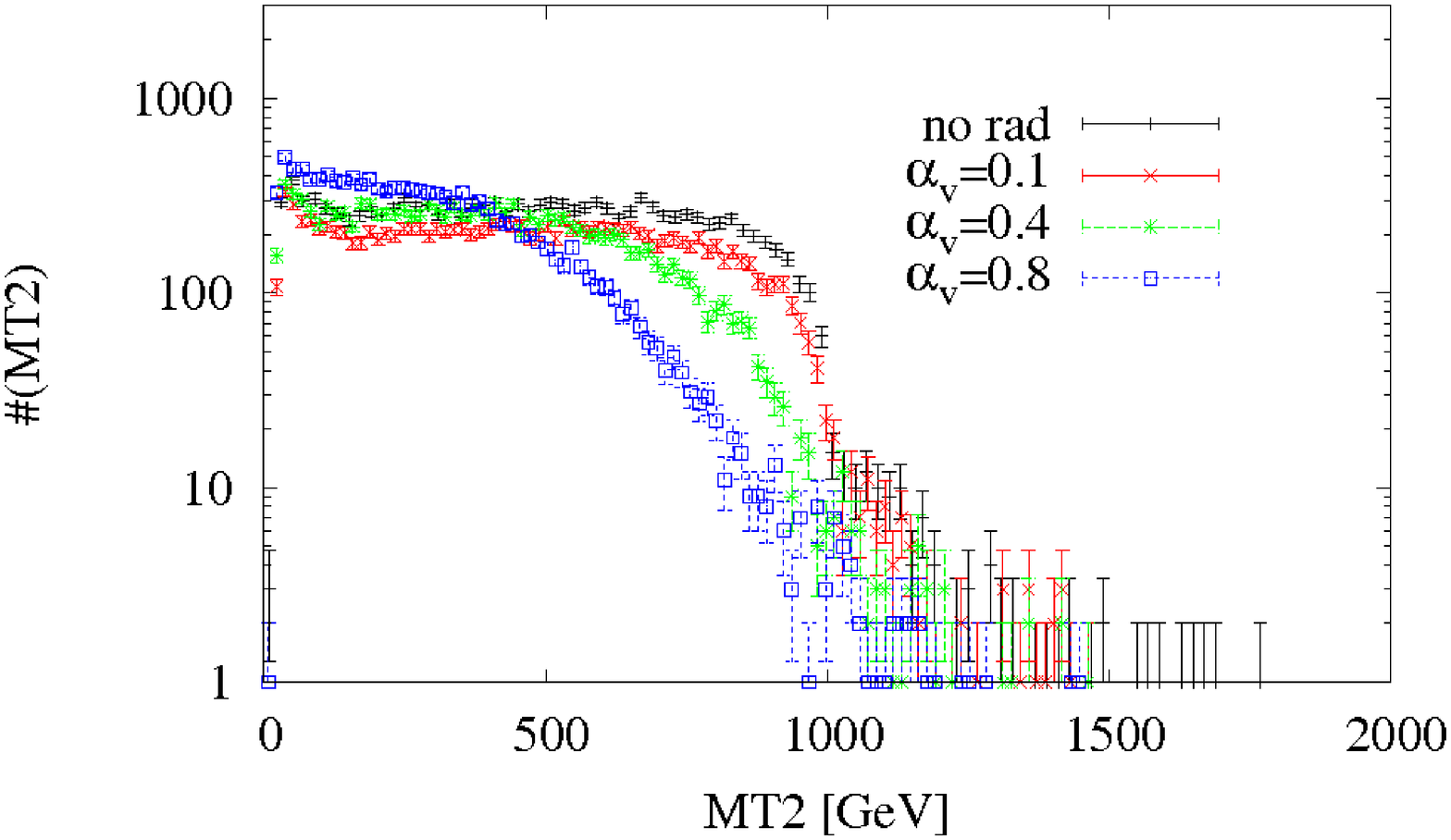,scale=0.4, angle=270}
\caption{MT2 distribution function for the communicator masses $M_{D_v}=1$ TeV when there is valley radiation $\alpha_v=$ 0.1 (red), 0.4 (green), 0.8 (blue) and when there is not (black curve). $\sqrt{s}=14$ TeV, $M_{q_v}=10$ GeV. Number of events per 15 GeV bin.}
\label{fig:MT2_LHC_14}
}

Notice how in this case one can actually separate (at least before any background or detector simulation is taken into account) the curve with $\alpha_v=0.1$  and the one without the radiation. 

In the above study the assumption is that events are generated with a bare mass $M_{q_v}$ and analyzed with the same (or almost the same) trial mass $\mu_{q_v}=M_{q_v}$. 
From the discussion in subsection \ref{subsec:MT2}, we could conclude that it is not very likely that we would know the mass of the $q_v$  with high precision  when valley radiation is present. 

%where we discuss the MT2 mqv dependence
In a hidden valley scenario the mass of the $q_v$ is assumed to be much
lighter than the $D_v$ mass, so we do not expect the MT2 to be very sensitive
to $q_v$ mass differences of the order $\Delta M_{q_v}\ll M_{D_v}$. Indeed, the
authors of \cite{Cho:2007qv} and \cite{Barr:2010zj} show that the MT2$^{\max}$
mass dependence on the trial mass $\mu_{q_v}$ is rather weak so long as $\mu_{q_v}<M_{q_v}$, whereas MT2$^{\max}$ grows much more rapidly when $\mu_{q_v}>M_{q_v}$. The $\mu_{q_v}<M_{q_v}$ case  is exactly what one observes in Fig. \ref{fig:MT2_muqv}. 

\FIGURE[ht]{
\epsfig{file=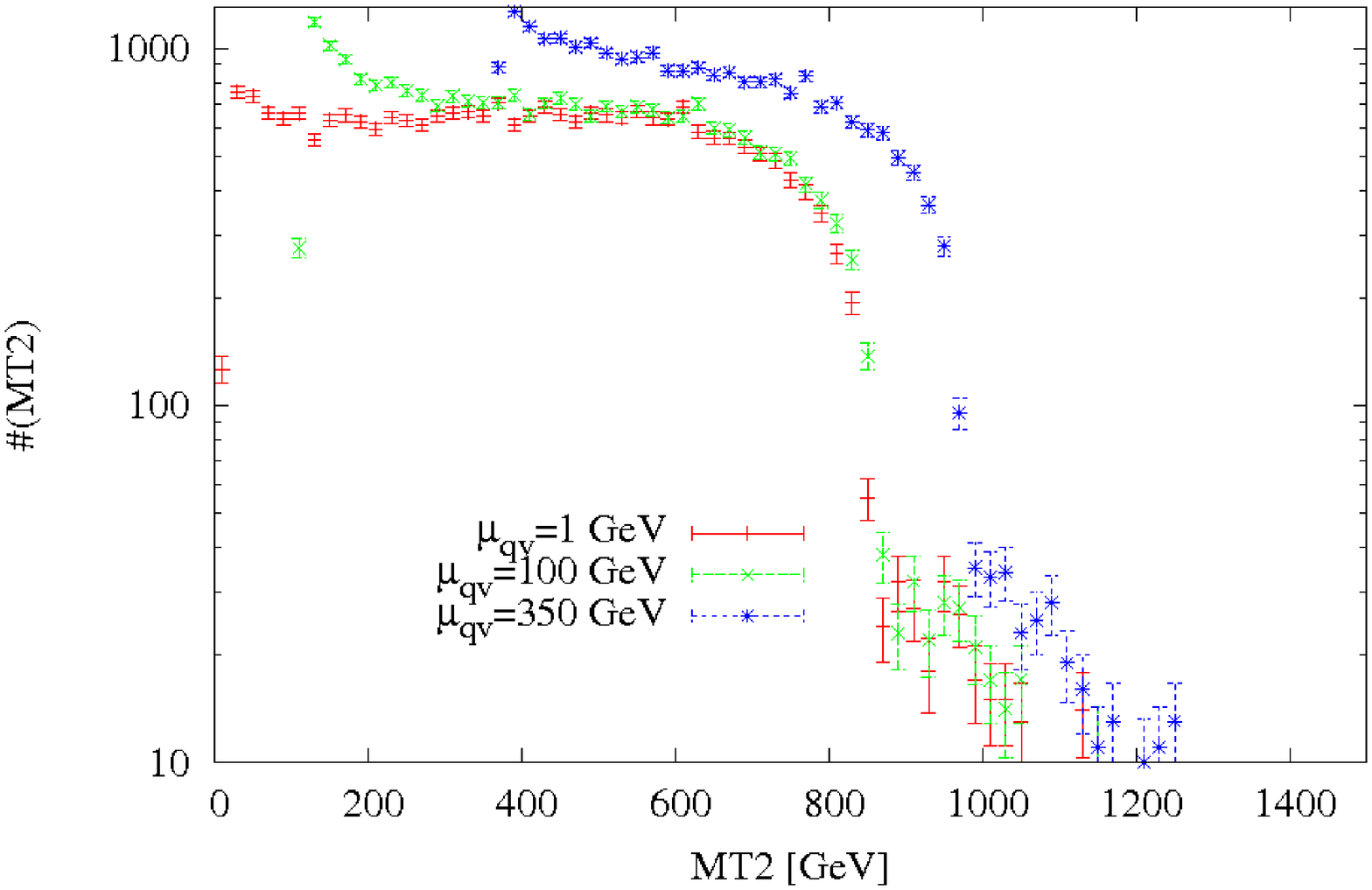,width=.31\textwidth,angle=270}
\epsfig{file=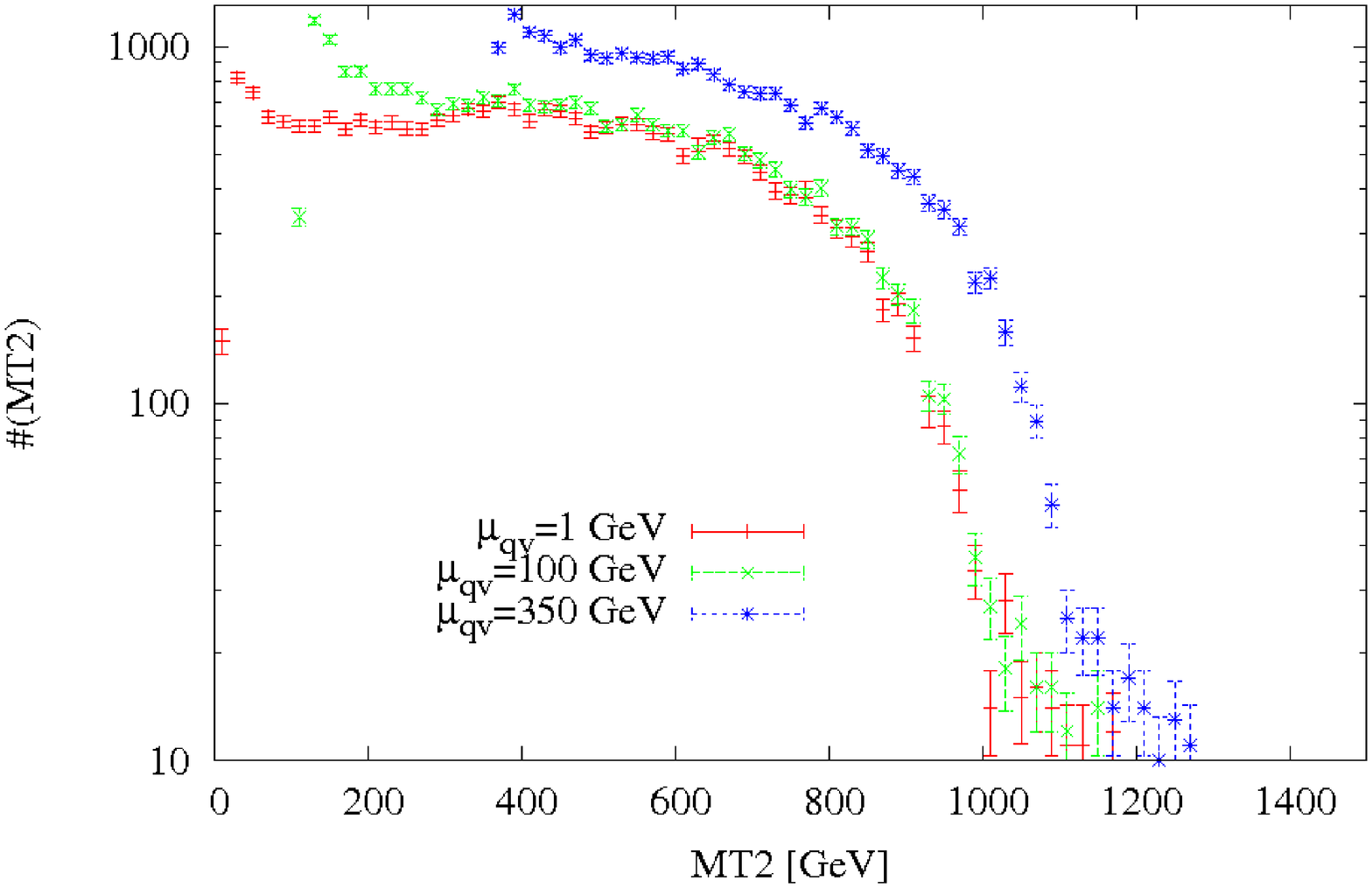,width=.31\textwidth,angle=270}
\caption{Left: model a) comparison  between different $\mu_{q_v}=1, 100, 350$ GeV for $M_{q_v}=395$ GeV. Right: model b) comparison  between different $\mu_{q_v}$ for $M_{q_v}=10$ GeV, $\alpha_v=0.28$, $\langle M^{\mathrm{eff}}_{q_v}\rangle=395$ GeV. In both models $M_{D_v}=1$ TeV, $\sqrt{s}=14$ TeV and  the luminosity is assumed to be $L=100$ fb$^{-1}$, 20 GeV bins.}
\label{fig:MT2_muqv}
}

Let us however make the conservative assumption that we only know the order of magnitude of the $M_{q_v}$. 
Imagine trying to distinguish between the two models we described in Fig. \ref{fig:models},  a) the model with no radiation and b) the model with the radiation, when $\langle M^{b,\mathrm{eff}}_{q_v}\rangle=M^b_{q_v}$ .  To be more concrete, assume a) has a  fixed value mass $M^a_{q_v}=395$ GeV and b) has an invisible particle mass  $M^b_{q_v}=10$ GeV and  $\alpha_v=0.28$.  We choose the  value $\alpha_v=0.28$ so that $\langle M^{b,\mathrm{eff}}_{q_v}\rangle=M^a_{q_v}$.

In Fig. \ref{fig:MT2_muqv}, left side, we see the MT2 distribution for model a) for the case $M_{D_v}=1$ TeV. As one may observe, this is a function of the trial mass $\mu_{q_v}$, the best profile being the one with $\mu_{q_v}=M^â_{q_v}$. For $\mu_{q_v} \ll M^a_{q_v}$, even for substantial changes in $\mu_{q_v}$ the MT2$^{\max}$ does not change much. This essentially confirms what was reported by \cite{Cho:2007qv,Barr:2010zj}.  

On the right side of Fig. \ref{fig:MT2_muqv} one may see the same distributions for the model with radiation. Contrary to what one would expect from the na\"ive arguments given in the previous subsection, we see that choosing $\mu_{q_v}\sim \langle M^{b,\mathrm{eff}}_{q_v}\rangle$ does not give the best description of the system. The MT2 curve overshoots the $M_{D_v}$ value by a good 10\%. 
As anticipated in the previous subsection, this is due to the invariant mass
distribution spread. Looking at  Fig.~\ref{fig:invmass_MT2}, one may see that
the invariant mass distribution has a wide spread, so event by event there
could be large variations in the $M^{b,\mathrm{eff}}_{q_v}$. If one chooses a
$\mu_{q_v} \ll \langle M^{b,\mathrm{eff}}_{q_v}\rangle$ to analyze the set of
events, for example the $\mu_{q_v}=1$ GeV chosen in Figure \ref{fig:MT2_muqv},
most of the events will have a "real'' $q_v$ mass, the invariant mass
$M^{b,\mathrm{eff}}_{q_v}$, larger than the trial mass $\mu_{q_v}$. Since
MT2$^{\max}(\mu_{q_v})<$ MT2$^{\max}(M^{b,\mathrm{eff}}_{q_v})$ when $\mu_{q_v}
< \langle M^{b,\mathrm{eff}}_{q_v}\rangle$,   these points do not contribute
to increase the MT2$^{\max}$ value much, and the distribution will resemble rather closely the one obtains for 
$M^b_{q_v}=10$ GeV, apart from the softening in the shoulder. This is precisely what happens in the $\mu_{q_v}=1,100$ GeV curves in Fig. \ref{fig:MT2_muqv}.  

When  one takes a $\mu_{q_v} \sim \langle M^{b,\mathrm{eff}}_{q_v}\rangle$ instead, e.g. $\mu_{q_v}=350$ GeV (which according to the na\"{i}ve arguments of the last subsection should have been the best of the three $\mu_{q_v}$ choices), more and more events have a "real'' invisible particle mass $M^{b,\mathrm{eff}}_{q_v}<\mu_{q_v}$. In   Fig.  ~\ref{fig:invmass_MT2} these are the points to the left of the  $\mu_{q_v}=350$ GeV line. These events will, if there is enough statistics, give  MT2$^{\max}(\mu_{q_v})>$MT2$^{\max}(M^{b,\mathrm{eff}}_{q_v})$, the real $M_{D_v}$ value. This is what happens to the curve for $\mu_{q_v}=350$ GeV.

\DOUBLEFIGURE[t]
{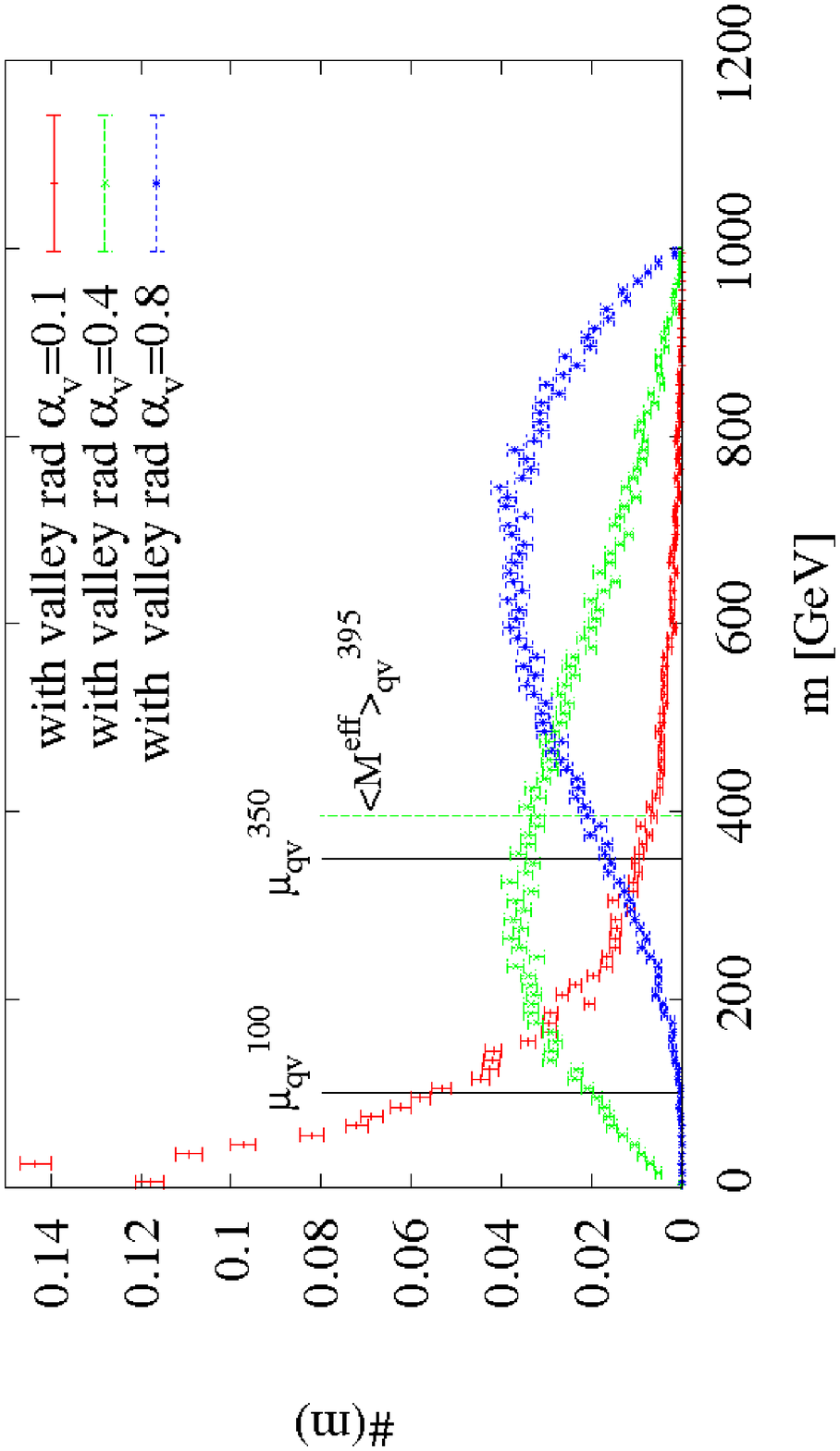,width=0.31\textwidth,angle=270}
{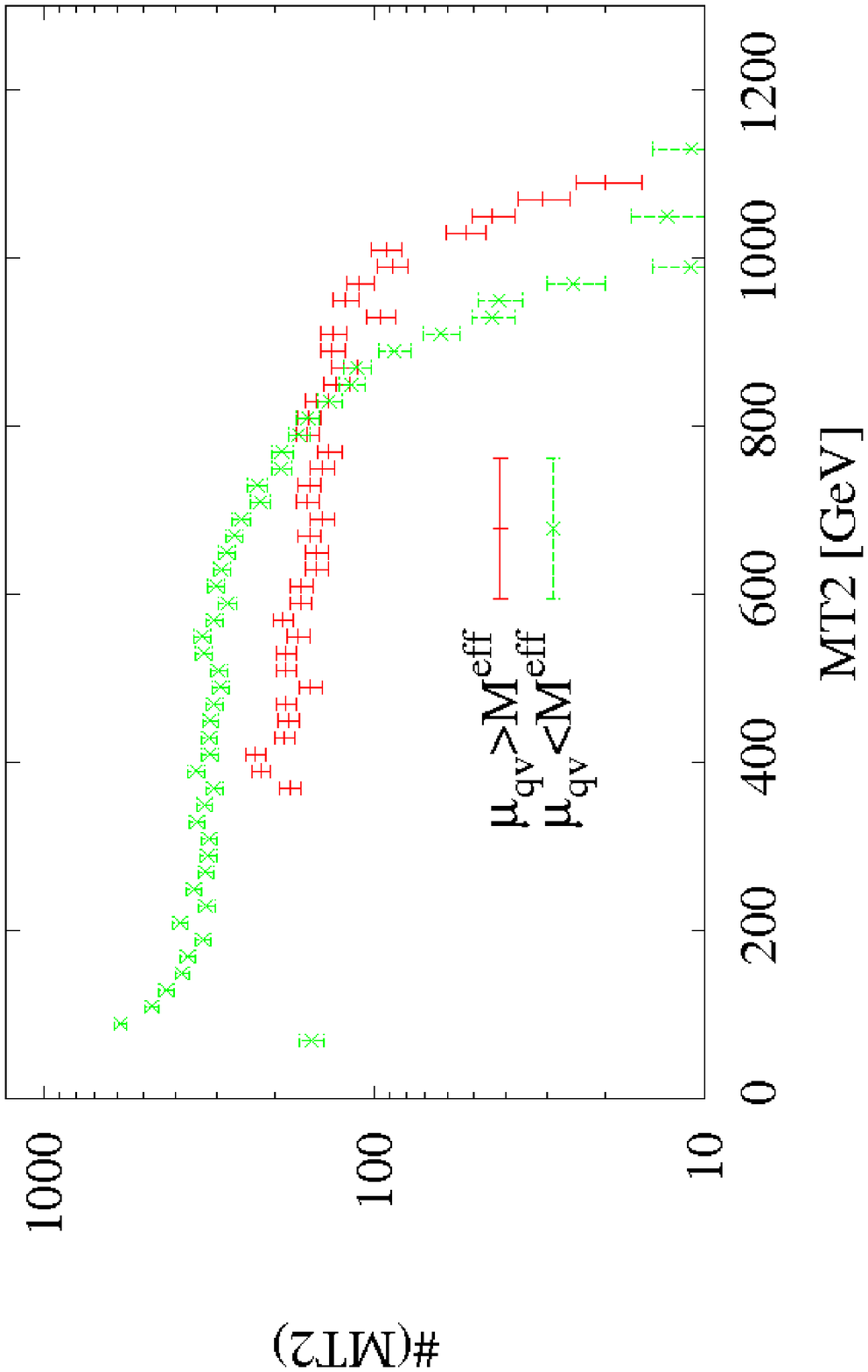,width=0.31\textwidth,angle=270}
{The invariant mass distribution for the $q_v$ and the trial masses we considered, $\mu_{q_v}=1,100,350$ GeV.\label{fig:invmass_MT2}}
{MT2 distribution when both $q_v$  invariant masses  $M^{b,\mathrm{eff}}_{q_v}<\mu_{q_v}$ (upper) and when both   are  $M^{b,\mathrm{eff}}_{q_v}>\mu_{q_v}$ (bottom), when $M^b_{q_v}=10$ GeV, $\alpha_v=0.28   $ and $\mu_{q_v}=350$ GeV.\label{fig:MT2_muqv2}}

To prove this point, we separately plot the MT2 distributions for the events
with $M^{b,\mathrm{eff}}_{q_v}<\mu_{q_v}$ and the ones with $M^{b,\mathrm{eff}}_{q_v}>\mu_{q_v}$. As Fig. \ref{fig:MT2_muqv2} shows, for the former set MT2$^{\max}\ge M_{D_v}$.

\FIGURE[ht]{
\epsfig{file=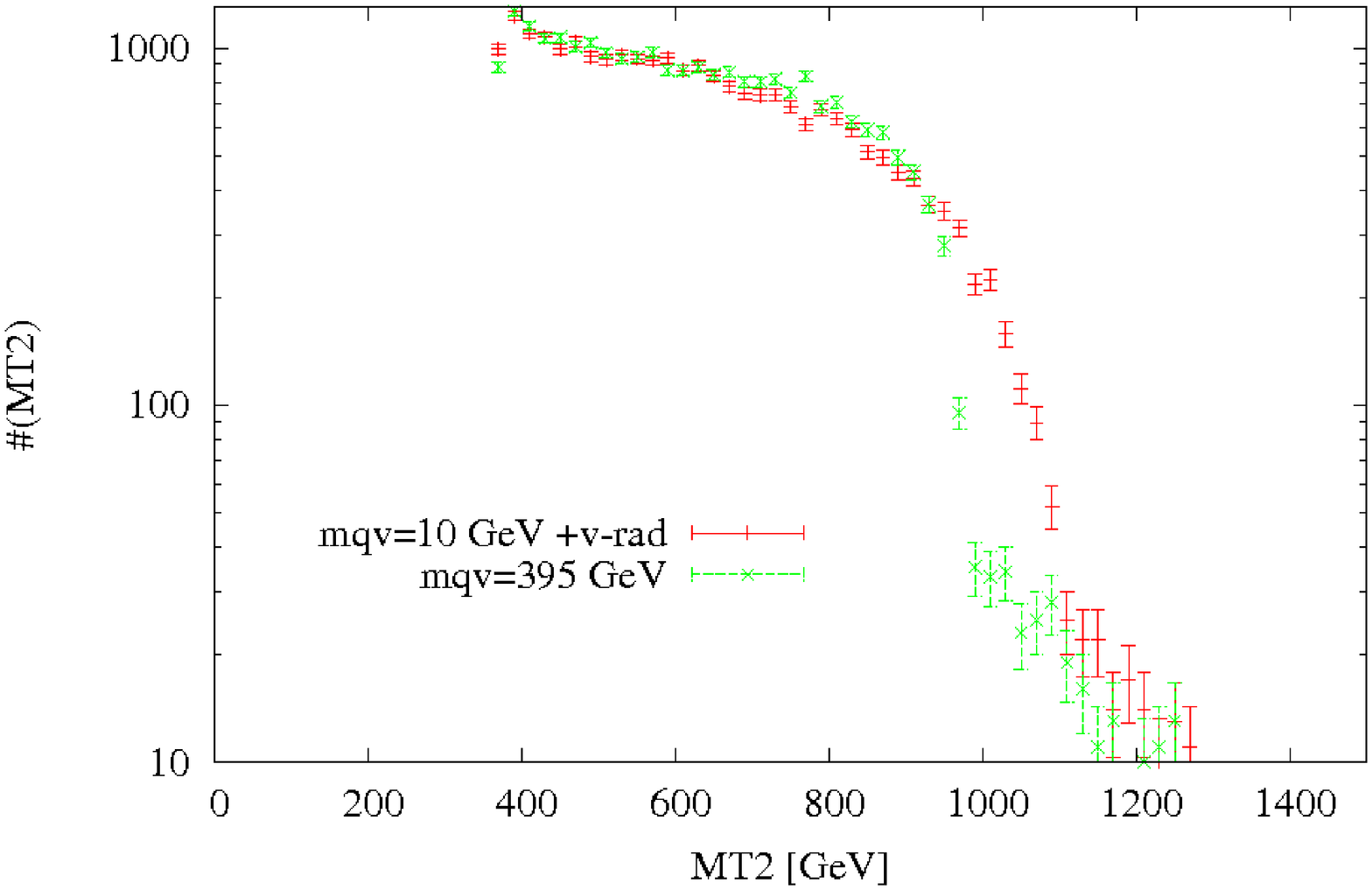,width=0.4\textwidth,angle=270}
\caption{Comparison  between model a) $M_{q_v}=395$ GeV  and  b) $M^b_{q_v}=10$ GeV, $\alpha_v=0.28$, $\langle M^{b,\mathrm{eff}}_{q_v}\rangle=395$ GeV for $\mu_{q_v}=350$ GeV. In both models $M_{D_v}=1$ TeV, $\sqrt{s}=14$ TeV and  the luminosity is assumed to be $L=100$ fb$^{-1}$, 20 GeV bins. Notice that in this case the curve with the radiation is the lower one.}
\label{fig:MT2_IMvsFix}
}

This said, we may return to the issue of distinguishing between the two models a) and b). As one may see in Figure \ref{fig:MT2_IMvsFix}, the two curves corresponding to the two models have very different shapes, the one with no radiation being the sharper one. Notice however that now the curve with radiation lies above the one without. This is not in contradiction with what we have shown in the previous sections.

%in case one has no clue about the invisible particle mass 

Summarizing, 
in the not-so-strong coupling regime $\alpha_v \sim 0.1$, when the invariant mass distribution is strongly peaked in $M_{q_v}$, one may\footnote{At least before background and detector analysis.} distinguish two models with  the same values for $M_{D_v}$ and $M_{q_v}$, one with v-radiation and one without. The curve corresponding to the no hidden radiation model will have a steeper drop in the shoulder region.

For larger couplings, $\alpha_v>0.2$, one should note that two MT2 distributions with the same endpoints may correspond to different $M_{D_v}$ values in the two cases, depending on whether the majority of the events have an invariant mass which is greater or smaller than the  trial mass. A very conservative approach to solving this problem could simply be to assume $\mu_{q_v}=0$ GeV or in any case $\mu_{q_v}\ll \langle M^{\mathrm{eff}}_{q_v}\rangle$.   

\section{Conclusions}
We have here addressed the issue of detecting and identifying hidden radiation through its influence on SM parton showers, and in general through its impact on visible particle kinematic distributions.  We  have done so in the context of  Hidden Valley models, which we find are well suited to display the effects, but the specific models studied should be viewed only as representatives of a broad range of possible models with new symmetries.  
Thus, while we focus on the phenomenology of a fairly generic toy model, we  also provide  tools in  \textsc{Pythia~8} to simulate the effects of hidden radiation in various other hidden valley scenarios, e.g. different gauge groups, particle contents, and gauge and decay couplings. The novel feature in these tools is the interleaved SM and valley parton shower, i.e. the competition between visible and hidden radiation. 

Our preliminary  study of the  phenomenology of the toy model at $e^+ e^-$ and at LHC colliders shows the following. 

%at ILC
At an 800 GeV ILC collider we could expect to observe hidden radiation for valley gauge couplings as small as $\alpha_v\ge 0.05$, so long as the mass of the communicator is smaller than 300 GeV and $M_{q_v}\ll M_{E_v}$. 

%LHC phenomenology
For the LHC phenomenology we need to distinguish between the first two year running with $\sqrt{s}=7$ TeV and $L=1$ fb$^{-1}$  and later years with full design energy $\sqrt{s}=14$ TeV and $L=100$ fb$^{-1}$. Whether one could observe hidden radiation or not depends strongly on the communicator mass, which determines the amount of statistics.
The main signal of the hidden radiation --- at least in our studies --- is the softening of the shoulder of the MT2 distribution.
%at LHC 7 TeV run 
In the lower-energy case, for communicator masses  around 500 GeV or higher, the statistics is so poor that one cannot expect to distinguish even a very strong coupling $\alpha_v=0.4$. For two models with and without hidden radiation, with equal $M_{D_v}$ masses in the [300, 400] GeV range and equal $M_{q_v}\le 50$ GeV, one would need a valley coupling of the order $\alpha_v\ge 0.2$ or larger to induce large enough effects on the MT2 distribution to distinguish between the two.
%LHC @14 TeV 
In the higher-energy case, for order TeV communicator masses and $M_{q_v}$ smaller than 100 GeV, the MT2 distributions show sizable changes already for  $\alpha_v= 0.1$.

%mu_qv
We have also studied how the MT2 distribution depends upon the $M_{q_v}$, the invariant mass $ M^{\mathrm{eff}}_{q_v}$, and the trial mass $\mu_v$. In the case of a Hidden Valley scenario, 
$M_{q_v}$ is always assumed to be $M_{q_v}\ll M_{D_v}$. Taking a trial mass 
$\mu_{q_v}\ll M_{D_v}$ as an input parameter for the MT2 is thus a natural choice. 
When the $q_v$ mass is larger though, e.g. $M_{q_v}\sim M_{D_v}$, the issue of the trial mass 
$\mu_{q_v}$ is no longer so trivial. 
The $ M^{\mathrm{eff}}_{q_v}$ can be a broad distribution when 
radiation is present. Especially  in the strong interaction case, $\alpha_v\ge 0.2$, $\langle M^{\mathrm{eff}} \rangle$ is strongly shifted towards $M_{D_v}$. This means that when one chooses a trial mass $\mu_{q_v} \sim \langle M^{\mathrm{eff}} \rangle$, roughly half of the events will have $\mu_{q_v} >M^{\mathrm{eff}}$, causing the MT2$^{\max}$ to overshoot the real value. In this case the new masses thus have to be extracted from a combined fit, in which both masses and couplings enter as unknowns.
 
Further studies of the background and detector simulations should follow, both for the kind of scenarios we have explored here and for other possible ones.
 This preliminary study,  however, shows unequivocably that parton showers are a key tool in determining the presence of new hidden gauge groups and in the exploration of the hidden sector gauge group dynamics.

\acknowledgments

We would like to thank Peter Skands for suggesting this study and for the profitable exchanges we had throughout the whole project. We also acknowledge helpful
discussions with Matt Strassler.

\end{document}